\documentclass[a4paper,amsmath,amssymb]{jpconf}

\usepackage{graphicx}
\usepackage{amssymb}

\usepackage{xcolor}
\usepackage{topcapt}

\newcommand{\be}{\begin{equation}}
\newcommand{\ee}{\end{equation}}
\newcommand{\ba}{\begin{eqnarray}}
\newcommand{\ea}{\end{eqnarray}}
\newcommand{\ci}[1]{\cite{#1}}
\newcommand{\lab}[1]{\label{#1}}
\newcommand{\beq}{\begin{equation}}
\newcommand{\eeq}{\end{equation}}
\newcommand{\beqs}{\begin{eqnarray}}
\newcommand{\eeqs}{\end{eqnarray}}

\newfont{\prg}{cmsy10}

\newcommand{\dss}{\Delta \sigma^{s} }
\newcommand{\dws}{\Delta \sigma^{d}}
\newcommand{\gev}{\,{\rm GeV}}
\newcommand{\tev}{\,{\rm TeV}}
\newcommand{\gvs}{\,{\rm (GeV}/c)^{2}}

\begin{document}
\title{Nucleon structure and spin effects in elastic hadron scattering  }

\author{O.V. Selyugin}

\address{ BLTPh, JINR,  Dubna, Russia}

\ead{selugin@theor.jinr.ru}

\begin{abstract}
 Soft diffraction phenomena in  elastic nucleon scattering
 are considered from the viewpoint of  the  spin dependence of the interaction potential.
Spin-dependent pomeron effects are analyzed for elastic $pp$
scattering,   spin-dependent differential cross sections and
spin correlation parameters are  calculated. 
  The spin correlation parameter $A_N$  is examined
 on the basis of experimental data from $\sqrt{s} = 4.9 \ GeV$ up to $23.4 \ $GeV
 in the framework of the extended High Energy Generalized Structure (HEGS) model.
   It is shown that the existing experimental data of  proton-proton and proton-antiproton
 elastic scattering at high energy in the region of the diffraction minimum and at
  large momentum transfer give the support of the existence of the energy-independent part
  of the hadron spin flip amplitude.
\end{abstract}




%

 \section{Introduction}

      Determination of the structure of the hadron scattering amplitude
  is an important task for both  theory and experiment.
  Perturbative Quantum Chromodynamics cannot be used
  in calculation of the real and imaginary
  parts of the scattering amplitude in the diffraction range.
 A worse situation holds for  spin-flip parts of the scattering
 amplitude in the domain of small transfer momenta. On the one hand,
 the usual representation tells us that the spin-flip amplitude  dies
 at superhigh energies, and on the other hand, we have  different
 nonperturbative models which lead to a nondyig spin-flip amplitude
 at superhigh energies \ci{bsw,zpc,ans,ETT-79}.

    The researches into  the spin-dependent structure of the hadron scattering amplitude
    are important for various  tasks. On the  one hand, the spin amplitudes
    constitute the spin portrait of the nucleon. Without
    knowing  their energy and momentum transfer dependence, it is  impossible
    to understand spin observable of nucleon scattering of nuclei.
    Such knowledge is also needed for the studies of very subtle effects,
     such as some attempts
    to search for a null-test signal of T-invariance violation under P-parity conservation
    in a $pd$ double polarization collision at  SPD NICA energies \cite{Uzikov}.
    Parity violation in the interaction of longitudinally polarized protons or deuterons
    with an  unpolarized target have been discussed in \cite{NNN1},
     and the estimates of the P-odd asymmetry in nucleon-nucleon scattering in the NICA energy range
      were reported in \cite{NNN2}.
    This is especially important for  future fixed-target experiments at LHC,
    in which $pp$, $pd$ and $pA$ collisions can be performed at
    $\sqrt{s_{NN}} =115$~GeV as well as $Pbp$ and $PbA$ collisions at  $\sqrt{s_{NN}} =72$~GeV.

  The study  of elastic  scattering requires  knowledge
of  the  properties of the pomeron,
the object determining  the interaction of hadrons in elastic and exclusive processes. 
In this case,  the study of the structure  and  spin  properties of
both the hadron  and the pomeron  acquires  a  special  role \cite{lap}.
The vacuum $t$-channel amplitude is usually associated with
two-gluon exchange in QCD \cite{low}. The properties of the spinless pomeron
 were analyzed on the basis of a QCD model,
by taking into account the non-perturbative properties of the theory
\cite{la-na,don-la}.
Now we recognize that the research into
the pomeron exchange
requires not only a pure elastic process but also many
physical processes involving electroweak boson exchanges \cite{Jenk1}.
There are two approaches to the pomeron, the "soft" pomeron built of
multiperipheral hadron exchanges and a more current perturbative-QCD
"hard" pomeron built of the gluon-ladder.

The spin structure of the pomeron is still an unresolved question
in  diffractive scattering of polarized particles.
There have been many observations of spin effects at high energies and
at fixed momentum transfers \cite{krish,nur}; several
attempts to extract the spin-flip amplitude from the experimental
data  show that the ratio of spin-flip to spin-nonflip
amplitudes can be non-negligible and may be independent of energy
\cite{akch,sel-pl}.


It is generally believed, based on calculations of the simplest QCD diagrams,
that the spin effects decrease as an inverse power of the center-of-mass energy and
that the pomeron exchange does not lead to appreciable spin effects in
the diffraction region at super-high energies.
Complete calculations of the full set of helicity scattering amplitudes
in the diffraction region cannot be carried out presently since they
require extensive treatment of confinement and contributions from many
diagrams.  Semi-phenomenological models, however, have been developed
with parameters which are expected to be fixed with the aid of data from
experiments.      
 There is a specific features of model approaches to the description of  spin correlations parameters.
  They can be divided into two classes - low energy and high energy.
  In the region of low energies there is a wide range of experimental data. To describe the data quantitatively the models used  pure phenomenological approaches
  like \cite{Wakaizumi}  or obtained some qualitative descriptions of the data in some theoretical approaches
  \cite{Sibirt}. Above $\sqrt{s} = 10$ GeV the number of experimental data essentially decreases,
  practically gives only the values of the analysing power $A_N(s,t)$. The description of such data was presented
  in the old famous Bourrelly-Soffer model \cite{bsw} and  recently in the model with a huge number of  free
  parameters \cite{Martynov}. However, such models do not take into account in the analysis the data at $\sqrt{s} < 13$ GeV. 
%



Some models predict non-zero spin effects as $s \to
\infty$ and $|t|/s \to 0$.
   In these studies, the spin-flip amplitudes, which lead to weakly altered
spin effects with increasing energy, are connected with the structure of
hadrons and their interactions at large distances \cite{soff,zpc}. In
 \cite{soff}, the spin-dependence of the pomeron term is
constructed within model rotation of matter inside the proton. This approach is
based on Chou and Yang's concept of hadronic current density \cite{c-y}.
This picture can be related with the spin effects determined
by higher-order  $\alpha_{s}$ contributions in the framework of PQCD.

The high energy two-particle amplitude determined by pomeron
exchange can be written in the form:

\begin{equation}
T(s,t)=is \ I\hspace{-1.mm}P(s,t) V_{h_1h_1 I\hspace{-0.8mm}P}^{\mu} \otimes
V^{h_2h_2sp I\hspace{-0.8mm}P}_{\mu}.    \label{tpom}
\end{equation} 
%
Here 
  $   I\hspace{-1.mm}P(s,t)$  
 is a function caused by a pomeron
with a weak energy dependence $\sim (\ln{s})^n$  and
$V_{\mu}^{hhI\hspace{-0.8mm}P}$ are the pomeron-hadron vertices.
The perturbative calculation of the pomeron coupling structure
is rather difficult, and the non-perturbative contributions are important
for momentum transfers of a few $\gvs$.
The situation  changes dramatically when large-distance loop
contributions are considered, which leads to a more complicated spin structure
of the pomeron coupling.
 As a result,
spin asymmetries appear that have a weak energy dependence as $s \to
\infty$. Additional spin-flip contributions to the quark-pomeron vertex
may also have their origins in instantons, {\it e.g.} \cite{do,ans}.
 Note that in the framework of the perturbative QCD, the analyzing power of hadron-hadron
scattering was shown to be
of the order:
$$  A_N \ \propto \ m \alpha_s / \sqrt{p_{t}^{2}}$$
where $m$ is around hadron mass \cite{ter}.  Hence, one would
expect a large analyzing power for moderate $p_{t}^{2}$,
where the spin-flip amplitudes are expected to be
important for diffractive processes.


  Now there are many different models to  describe  the elastic
  hadron scattering amplitude at small angles (see reviews
  \cite{Rev-LHC,Paccomomi,Martynov}).
  They lead to  different predictions of the structure of the scattering
  amplitude at super-high energies.
  The diffraction  processes at  very high
 energies, especially at
    LHC energies  are not simple asymptotically
     but can display complicated features    \cite{FS-07,dif04}.
    Note that the interference of  hadronic
 and electromagnetic amplitudes can give an important contribution
  not  only at very small transfer momenta but also in the range of the
 diffraction minimum \cite{lap}.  However, one should also know the
 phase of the interference of the Coulomb and hadron amplitudes at
 sufficiently large transfer momenta
 and  the contribution of the hadron-spin-flip amplitude to the CNI effect
  \ci{trosh,soff}.
     Now we cannot exactly calculate all
  contributions and find their energy   dependences. But a great amount of
  the experimental material at low energies allows us to make complete
  phenomenological analyses and find the size and form of
  different parts of the hadron scattering amplitude.  The difficulty
  is that we do not know the energy dependence of these amplitudes and
  individual contributions of the asymptotic non-dying spin-flip
  amplitudes.

 From a modern point of view, the structure of a hadron can be described by
the Generalized parton distribution (GPD) functions \cite{{Muller,Ji97,R97}}
combining our knowledge about the one-dimensional parton
distribution in the longitudinal momentum with the
impact-parameter or transverse distribution of matter in a hadron
or nucleus. They allow one to obtain a 3-dimensional picture of the nucleon (nucleus)
\cite{Burk1,Burk2,Diehl}.

     In the general picture,
    hadron-hadron processes determined by the strong interaction with
 the  hadron spin equal to $1/2$ can be represented by some combinations of three vectors
   in the  
    center of mass system. 
   In the c.m.s. there are only two independent three-dimensional momenta.
   In the case of the elastic scattering,
     $\vec{p}_1 = - \vec{p}_2$ and $\vec{p}_3 = - \vec{p}_4$
   Using the initial and final momenta  ${\bf p}$ and $p^{\prime}$
   and their unity vectors ${\bf \hat{p}}$ and  ${\bf \hat{p}^{\prime}}$ ,
   so that ${\bf \hat{p}}={\bf p}/{|\bf p}|$
   and ${\bf \hat{p}}={\bf p^{\prime}}/{|\bf p}^{\prime}|$,
     one can  obtain three independent combinations
$${\bf \hat{l}} \equiv \frac{ {\bf p + p^{\prime}} }{ |{\bf p + p^{\prime}}| }; \ \ \
  {\bf \hat{q}} \equiv \frac{ {\bf p - p\prime } }{ |{\bf p - p\prime }| }; \ \ \
  {\bf \hat{n}} \equiv \frac{ {\bf p \times p^{/}}}{|{\bf p \times p^{/}}| }.
$$
  The vectors ${\bf \hat{l}}$, ${\bf \hat{q}}$, ${\bf \hat{n}}$ and spin-vectors
  ${ \hat{ \sigma_{1}}}$ and ${\hat{\sigma_{2}}}$  create eight independent
  scalars \cite{Nelipa}
  $({ \hat{\sigma_{1}}}  {\bf \hat{n}})  ({ \hat{\sigma_{2}}}  {\bf \hat{n}}) \ \ \ $,
  ${ \hat{\sigma_{1}}} {\bf \hat{n}} + { \hat{\sigma_{2}}}  {\bf \hat{n}}  \ \ \ $,
${ \hat{ \sigma_{1}}}  {\bf \hat{n}} - { \hat{\sigma_{2}}}  {\bf \hat{n}} \ \ \  $,
$({ \hat{\sigma_{1}}}  {\bf \hat{I}})  ({ \hat{\sigma_{2}}}  {\bf \hat{I}}) \ \ \  $,
$({ \hat{\sigma_{1}}}  {\bf \hat{q}}) \times ({ \hat{\sigma_{2}}}  {\bf \hat{q}}) \ \ \  $,
$({ \hat{\sigma_{1}}}  {\bf \hat{q}})  ({\hat{\sigma_{2}}}  {\bf \hat{I}}) \ \ \  $,
$({ \hat{\sigma_{1}}}  {\bf \hat{I}})  ({ \hat{\sigma_{2}}}  {\bf \hat{q}})  \ \ \ $,
$[{ \hat{\sigma_{1}}}  { \hat{\sigma_{2}}}]   {\bf \hat{n}} $.

   The main experimental data show the conservation of  time parity,  
    charge conjugation, and space parity in the strong interaction processes.
Then,
  under  time inverse  ${\hat {\sigma_{1}}}$ 
  changes to 
   $-{ \hat{\sigma_{1}}}$, and
  ${\bf \hat{q}} \rightarrow {\bf \hat{q}}$,  ${\bf \hat{n}} \rightarrow - {\bf \hat{n}} \ \ \ $,
   ${\bf \hat{I}} \rightarrow - {\bf \hat{I}}$,  the combinations
   $[{ \hat{\sigma_{1}}}  { \hat{\sigma_{2}}}]   {\bf \hat{n}} \ \ \ $ and    $({ \hat{\sigma_{1}}}  {\bf \hat{q}})  ({ \hat{\sigma_{2}}}  {\bf \hat{I}}) \ \ \  $,  $({ \hat{\sigma_{1}}}  {\bf \hat{I}})  ({ \hat{\sigma_{2}}}  {\bf \hat{q}}) $
   have to be removed as a result of the time parity  conservation.
   If the interacting particles  are identical, as in the case of the  proton-proton elastic  scattering,
   their  combinations  should not be changed when replacing one particle by  another.
   As a result, the scattering amplitude is
\begin{eqnarray}
 \phi(s,t))=\phi_1(s,t)& +& \phi_2(s,t)
 ({ \sigma_1} \cdot {\bf{\hat{n}}})
		 ({\sigma_2} \cdot {\bf{\hat{n}}})
       + \phi_3(s,t)  ({\sigma_1} \cdot {\bf \hat{n}} + {\sigma_2} \cdot \bf{\hat{n}} )  \nonumber \\
      & + & \phi_4(s,t)({ \sigma_1} \cdot {\bf \hat{q}})
		 ({\sigma_2} \cdot {\bf \hat{q}})
       + \phi_5(s,t)({ \sigma_1} \cdot {\bf \hat{l}})
		 ({ \sigma_2} \cdot \bf{\hat{l}}),
\end{eqnarray}
 The amplitude 
 corresponds 
 to the spin-dependent interaction potential.
  It can be taken as a Born term of scattering  processes. The Born term of the amplitude in
 the  transfer momentum representation  can  be  obtained by the corresponding  amplitudes
  in the impact parameter representation
   \begin{eqnarray}
 \phi(s,b) \   =  -\frac{1}{2 \pi}
   \ \int \ d^2 q \ e^{i \vec{b} \cdot \vec{q} } \  \phi^{\rm Born}_{h}(s,q^2) \,,
 \label{tot02}
 \end{eqnarray}
 The corresponding amplitude is connected to the interaction potential  
in the position representation. 
  Using the standard Fourier transform \cite{Gold} of the potential $V(\vec{r})$,
 one can  obtain the Born term of the scattering amplitude.
   If the potentials $V(\vec{r}) = V_{1}(\vec{r}) + V_{5}(\vec{r})$ are
   assumed to have a Gaussian form
$$ V_{1,5}(\rho, z) \sim \ \int_{-\infty}^{\infty} e^{-B \ r^2} \ d z
 \ = \ \frac{\sqrt{\pi}}{\sqrt{B}} e^{-B \ \rho^2}, $$ 
  in the first Born approximation
  $\phi^{h}_{1}$ and $\hat{\phi_h}^{5}$
   will have  the same form
\ba
\phi^{h}_{1}(s,t) \sim  \int_{0}^{\infty} \ \rho \ d\rho
 \ J_{0}(\rho \Delta) e^{-B \ \rho^2} \ = \ e^{-B \Delta^{2}}, \lab{f1a}
\ea
\ba
\phi^{h}_{5}(s,t) \sim \int_{0}^{\infty} \ \rho^2 \ d\rho
 \ J_{1}(\rho \Delta) \  e^{\chi_{0}(s,\rho)}
   \ e^{-B \ \rho^2 } \ = \ q \ \ B \ e^{-B \Delta^{2}}  . \lab{f5a}
\ea In this special case, therefore,
 the  spin-flip and  ``residual''spin-non-flip amplitudes are
  indeed with the same slope  \cite{PS-Sl}.

  The first observation that the slopes do not coincide
 was made in \cite{Predazzi66}.
  It was found from the analysis of the
 $\pi^{\pm} p \rightarrow \ \pi^{\pm}p $ and
  $pp \rightarrow \ pp $
reactions
  at $p_L \ = \ 20 \div 30 \ GeV/c $
  that the slope of the ``residual'' spin-flip amplitude is about
   twice as large as
  the slope of the spin-non flip amplitude. This conclusion can also
  be obtained
  from the phenomenological analysis carried out in \cite{Wakaizumi} for
  spin correlation parameters of the elastic proton-proton scattering
  at $p_L \ = \ 6 \ GeV/c$.

   The model-dependent analysis based on all the existing experimental
   data of the spin-correlation parameters above $p_L \ \geq
 \  6 \ GeV$
   allows  us to determine the structure of the hadron spin-flip
   amplitude at high energies and to predict its behavior at
   superhigh energies \cite{z100}   
   This analysis shows
   that the ratios
   $Re \ \phi^{h}_{5}(s,t) / (\sqrt{|t|} \ Re \ \phi^{h}_{1}(s,t))$ and
   $Im \ \phi^{h}_{5}(s,t)/(\sqrt{|t|} \ Im \ \phi^{h}_{1}(s,t))$
   depend on $\ s$ and $t$. At small momentum transfers,
   it was found that the slope of the ``residual'' spin-flip
    amplitudes is approximately
   twice the slope of the spin-non flip amplitude \cite{JPS}.

  The electromagnetic current of a nucleon is
\ba
 J_{\mu} (P^{'},s^{'}; P,s)  = \bar{u} (P^{'},s^{'}) \Lambda_{\mu} (q,P) u(P,s)
 =  \bar{u} (P^{'},s^{'}) (\gamma_{\mu} F_{1}(q^2) +
   \frac{1}{2M} i \sigma_{\mu \nu }q_{\nu }F_{2}(q^{2}))u(P,s),
\ea
where $P,s,  (P^{'},s^{'}) $ are the four-momentum and polarisation of the incoming (outgoing) nucleon,
and $q = P^{'}- P $ is the momentum transfer.  The quantity  $  \Lambda_{\mu} (q,P) $ is the nucleon-photon vertex.

 If the potential has the spherical symmetry, the Born amplitude can be
 calculated as
  \ba
 \phi_{B}(t) = g/q \int_{0}^{\infty} r \ sin( q r )  V(\vec{r})  \ dr.
  \ea
 or in the impact parameter representation
    \ba
 \phi_{B}(b) = \frac{1}{4\Pi} \int_{0}^{\infty} r  V(\sqrt{(z^2+b^2)} ) 
  \ dr.
  \ea

  There are some different forms of the unitarization procedures \cite{Cud-Sel-nl,Cud-Pred-Sel-nl}.
   One of them is the standard eikonal representation, where
   the Born term of the scattering amplitude in the impact parameter representation
    takes  the eikonal phase and the total scattering amplitude is represented as
\begin{eqnarray}
\phi(s,t)=\frac{i s}{2 \pi} \int_{0}^{\infty} [1-\exp{(-\chi({\bf b })}]
	\exp{(-i {\bf{ q \cdot b}})} d^2 {\bf{ b}}.
\end{eqnarray}


 If the terms are taken into account to first order in the spin-dependent
  eikonals   of the spin-dependent eikonal amplitude, where
  the eikonal function $\chi(\bf{b})$ is a sum of the spin-independent
  central term  $\chi_{si}$, spin-orbit term - $\chi_{ls}$, and spin-spin
	 term $\chi_{ss}$,
 separate spin-dependent amplitudes  are written as follows:
\begin{eqnarray}
 \phi_{1s}(s,t)&=&i s \int b d b J_0(b q)[1-\exp{(-\chi_{si}(b))}]; \\
 \phi_{2s}(s,t)&=&i s \int b^2 d b J_1(b q) \exp{(-\chi_{si}(b))} \chi_{ls}(b); \\
 \phi_{3s}(s,t)&=& s \int b d b J_0(b q) \exp{(-\chi_{si}(b))}
                                                       \chi_{ss}(b).
\end{eqnarray}

  Using ordinary relations (see, for example, \cite{Lehar}), we can
 obtain  helicity amplitudes for small scattering angles and high
  energies:
\begin{eqnarray}
 \phi_1(s,t)&=&\phi_{1s}(s,t)-\phi_{2s}(s,t); \ \ \ \phi_2(s,t)=2 \phi_{2s}(s,t); \\ \nonumber
 \phi_3(s,t)&=&\phi_{1s}(s,t)+\phi_{2s}(s,t); \ \ \
 \phi_4(s,t)= 0; \ \ \  \phi_5(s,t)=i\phi_{3s}(s,t).
\end{eqnarray}

   The scattering amplitude of charged hadrons is represented as a sum of the
   hadronic and electromagnetic amplitudes $ \phi_{tot} = \phi^{em} + \phi^{h} $.
   The electromagnetic amplitude can be calculated in the framework of QED.
   In the one-photon approximation, we have \cite{bgl,nur1}
  \begin{eqnarray}
  \phi^{em}_1 &=& \alpha [f_{1}^{2}(t)(\frac{s-2 m^2}{t} + \frac{m^2}{2 p^2})
  - 2 f_{1} (t) f_{2}(t) -\frac{1}{2}f_{2}^{2}(t)(1-\frac{t}{4 p^2})];
   \\ \nonumber
  \phi^{em}_2 &=& \alpha [f_{1}^{2}(t)\frac{ m^2}{2 p^2}
  -f_{1}(t) f_{2}(t)
   + \frac{f_{2}^{2}(t)}{4 m^2}(s - 2 m^2 +\frac{t s}{8 p^2})];
   \\ \nonumber
  \phi^{em}_3 &=& \alpha (1 + \frac{t}{4 p^2}) [f_{1}^{2}(t) \frac{s-2 m^2}{t}
    + \frac{1}{2} f_{2}^2(t) ];
   \\ \nonumber
  \phi^{em}_{4} &=& - \phi^{em}_{2};
   \\ \nonumber
  \phi^{em}_5 &=& \alpha [-\frac{s (4p^2+t)}{t}]^{1/2}
        [f_{1}^{2}(t)\frac{ m}{4 p^2}
  - \frac{1}{2 m} f_{1}(t) f_{2}(t) + \frac{t}{16 m p^2} f_{2}^{2}(t)],
  \end{eqnarray}
  where $\alpha$ is the electromagnetic coupling constant, and
   $f_{1} (t)$ and $f_{2} (t)$ are the Dirac an Pauli  form factors of the proton
  \begin{eqnarray}
  f_{1}(t) = \frac{4 m^2 -t (1+k)}{4 m^2 -t} G_{d}
 \ \ \ \ f_{2}(t) = \frac{4 m^2 k}{4 m^2 -t} G_{d}
  \end{eqnarray}
  with
  $  G_{d} = (1- t/0.71)^{-2}$  
  (t is in GeV$^2$),  
   and $k= 1.793$ is the anomalous magnetic moment of the proton.

  In the high energy approximation, we obtain:
  \begin{eqnarray}
  \phi^{em}_1 = \alpha f_{1}^{2} \frac{s-2 m^2}{t},
\ \ \  \phi^{em}_3 = \phi^{em}_1,
  \ \ \ \phi^{em}_2 = \alpha  \frac{f_{2}^{2}(t)}{4 m^2} s,
 \\ \nonumber
  \phi^{em}_{4} = - \phi^{em}_{2}, \ \ \ \
  \phi^{em}_5 = \alpha \frac{s }{2m \sqrt{|t|}} f_{1}^{2},
  \end{eqnarray}
 The total helicity amplitudes can be written
 as
 $ \phi_i(s,t) = \phi^i_{N}(s,t) + \phi^i_{em}(t) \exp{i \alpha \varphi(s,t)}$  
  \ci{bgl}.

  The differential cross
  sections and spin correlation parameters are
\begin{eqnarray}
  \frac{d\sigma}{dt} =
 \frac{2 \pi}{s^{2}} (|\phi_{1}|^{2} +|\phi_{2}|^{2} +|\phi_{3}|^{2}
  +|\phi_{4}|^{2}
  +4 | \phi_{5}|^{2} ).	\label{dsdt}
\end{eqnarray}
\begin{eqnarray}
  A_N\frac{d\sigma}{dt}&=& -\frac{4\pi}{s^2} Im[(\phi_1+\phi_2+\phi_3-\phi_4) \ \phi_5^{*}).  \label{an}
\end{eqnarray}
and
\begin{eqnarray}
  A_{NN}\frac{d\sigma}{dt}&=& \frac{4\pi}{s^2} Re[(\phi_1 \phi_2^{*} - \phi_3 \phi_4^{*})
                + 2 |\phi_5|^{2}).  \label{ann}
\end{eqnarray}

\section{ 
  Coulomb 
  -nucleon phase factor}     

    In \cite{lap,bsw}, the importance of the CNI effects in the
     domain of  diffraction dip  
 was pointed out. In \cite{bsw}, the polarization at sufficiently
 low (now) energies with the CNI effect but without the phase
 of CNI and with a simple approximation
 of the hadron non-flip spin amplitude was calculated.
 However, the authors  showed for the first time that the CNI effect can be
 sufficiently large
 (up to $11\%$ at $p_{L} = 280 \ GeV/c$) in the region of non-small
 transfer momenta.

  The total amplitude including the electromagnetic and hadronic forces
  can be expressed as
\begin{eqnarray}
  F(s,t) =
  F_{C} \exp{(i \alpha \varphi (s,t))} + F_{N}(s,t),
\end{eqnarray}
and for the differential cross sections, neglecting the terms  proportional to $\alpha^2$, we have
\ba
d\sigma/dt&=&\pi [ (F_{C} (t))^2\!+\! ReF_{N}^{2}(s,t) +Im F_{N}^{2}(s,t)  \\
 &+&2 ( ReF_{N}(s,t) F_{C}(t) cos(\alpha \varphi(t)) + Im F_{N}(s,t) F_{C}(t) \ sin(\alpha \varphi(t)) )]
 \nonumber
\ea
with
\begin{eqnarray}
  \varphi(s,t) =  \varphi(t)_{C} - \varphi(s,t)_{CN} ,
\end{eqnarray}
 where   $\varphi(t)_{C}$ appears in the second Born approximation of
 the pure Coulomb amplitude, and the term $\varphi_{CN}$ is
 defined by the Coulomb-hadron interference.

   The  quantity $\varphi(s,t)$
 has been calculated and discussed  by many authors.
 For  high energies, the first results were obtained
 by  Akhiezer and  Pomeranchuk \cite{akhi}
 for the diffraction on a black nucleus.
 Using the WKB approach in potential theory,  Bethe  \cite{bethe}
 derived  $\varphi(s,t)$   for the proton-nucleus scattering.
  After some treatment improving this result \cite{rix}, the most
  important result was obtained by   Locher \cite{loch} and then by
  West and Yennie \cite{wy}.
   In the framework
  of the Feynman diagram technique in \cite{wy},
  a general expression  was obtained  for
   $\varphi_{CN}(s,t)$ in the case of pointlike particles
  in terms of the hadron elastic scattering amplitude:
 \begin{eqnarray}
  \varphi(s,t) = - \ln{(-t/s)} - \int^{S}_{0}
                  \frac{ d t^{'} }{ |t-t^{'}| } \  \
	       (1-\frac{  F_{N}(s,t^{'})  }
                       {  F_{N}(s,t)  } ).    \label{wy}
 \end{eqnarray}
   If  the hadron amplitude is chosen in  the standard Gaussian form \\
  $F_{N} = h \ \exp{(-B(s) q^{2}/2)}$, we can get
 \begin{eqnarray}
  \varphi(s,t) = \mp [\ln{(-B(s) t/2)} + \gamma],  \label{wyph}
 \end{eqnarray}
 where $-t=q^2$,  $B(s)/2$ is the slope of the nuclear amplitude,
  $\gamma$ is the Euler constant,
  and the upper (lower) sign
  corresponds to the scattering of particles with the same (opposite)
  charges.

 The impact of the spin of scattered particles was analyzed in
 \cite{lap,bgl} by using the eikonal approach for the scattering amplitude.
  Using the helicity formalism for high
  energy hadron scattering in \cite{bgl}, it was shown that at
  small angles, all the helicity amplitudes have the same $\varphi(s,t)$.
   The influence of the electromagnetic form factor
  of scattered particles on
  $\varphi_{C}$ and $\varphi_{CN}$
  in the framework of the eikonal approach was examined by Islam \cite{Islamphase}
  and with taking into account the hadron 
  form factor 
  in the simplest form
 by   Cahn \cite{can}.
  He derived for $t \rightarrow 0 $ the eikonal analogue (\ref{wy}) and
  obtained the corrections
\begin{eqnarray}
\varphi (s,t)&=&\mp [\gamma +\ln{ (B(s)|t| /2)}
         + \ln{ (1 + 8/(B(s)\Lambda ^2))} \nonumber\\
    & & + (4|t|/\Lambda ^2)\ \ln{ (4|t|/\Lambda^2)} + 2|t|/\Lambda^2],
                                                 \label{fit}
\end{eqnarray}
where
$\Lambda$ is a constant entering into the power dependent form factor. 
  The recent calculation of the phase factor was carried out in
  \cite{Petrovphase}.
   The calculations of the phase factor beyond the limit $t \rightarrow 0$
   were carried out in \cite{selmpl1,selmpl2,selphase}. As a result, for the total Coulomb scattering amplitude, we have the eikonal
approximation of the second order in $\alpha$
\begin{eqnarray}
F_c(q) = F_c^{1B} + F_c^{2B}
  = -\frac{\alpha}{q^2}
         [\frac{\Lambda^4}{(\Lambda^2+q^2)^2}]
  [1+i\alpha(\{ \ln(\frac{\lambda^2}{q^2})  \ + \ \nu_s \}],
\end{eqnarray}
where
\begin{eqnarray}
   \nu_s = A \ln(\frac{(\Lambda^2+q^2)^2}{\Lambda^2 q^2})
   + B  \ln(\frac{4 \Lambda^2}{(\sqrt{(4 \Lambda^2 +q^2}+q)^2}) \ +  \ C ,
                                                                 \label{e23}
\end{eqnarray}
 The coefficients $A,B,C$ are defined in \cite{selphase}.
     The numerical calculation shows
that at small $q^2$  the difference
between $\nu_s$ and $\nu_{c}$ is small, but
above $q^2=3.10^{-2} \ GeV^2$, it is rapidly growing.
It is clear that the solution of $\nu_{c}$ should be
bounded  at $-t= 3.10^{-2} \ GeV^2$.
  As a result, we have a sufficiently simple form of $\nu_s$ up to
  $|t| = 0.2 \ GeV^2$. It gives us
  the possibility to reproduce it by a simple phenomenological
  form that can be used in a practical  analysis of experimental data
  at small $t$:
  \begin{eqnarray}
     \nu_s \simeq = c_1 \log{(1 + c_{2}^{2} q^2)}  +4/q^{2},
  \end{eqnarray}
  where the constants $c_1$ and $c_2$ are defined by the fit $\nu_s$,
  $ c_1 = 0.11, \ \ \ \ c_2 = 20.$

  The total phase factor is
\begin{eqnarray}
 \varphi(s,t) = \ln{\frac{q^2}{4}} +2\gamma +\frac{1}{F_h(s,q)}
  \int_{0}^{\infty} \tilde{\chi}_{c}(\rho)
  (1 - \exp(\chi_h(\rho,s))J_{0}(\rho,q)d\rho , \label{fei2}
\end{eqnarray}
 with
\begin{eqnarray}
  \tilde{\chi}_c(\rho) = 2\rho \ln{\rho} +2\rho \biggl\{ K_{0}(\rho \Lambda)
  [1+ \frac{5}{24} \Lambda^2 \rho^2 ]
   +\frac{\Lambda \rho}{12} K_1(\rho \Lambda) [11+ \frac{1}{4} \Lambda^2 \rho^2] \biggr\}
\end{eqnarray}
  The calculated $\varphi(s,t)$, Eq.(\ref{fei2}), is an eikonal analog
 with taking account
 the hadron form factor
 of the expression  obtained by West and Yennie \ci{wy}
 from the Feynman diagram.

 \section{Nucleon form factors and GPDs}


There are various choices for the nucleon electromagnetic form
factors (ff), such as the Dirac and Pauli ff, $F_1^p(t),\ \
F_1^n(t)$ and $F_2^p(t),\ \ F_2^n(t),$ the Sachs electric and
magnetic ff, $G_E^p(t),\ \ G_E^n(t)$ and $G_M^p(t),\ \ G_M^n(t),$
\cite{Sach}.

The Dirac and Pauli form factors are obtained from a decomposition
of the matrix elements of the electromagnetic (e.m.) current in
linearly independent covariants made of four-momenta, $\gamma$
matrices and Dirac bispinors as follows:
$$<N|J_{\mu}^{e.m.}|N>=e\bar u(p')[\gamma_{\mu}F_1^N(t)+{i\over{2m}}\sigma_{\mu\nu}(p'-
p)_{\nu}F_2^N(t)]u(p),$$ where $m$ is the nucleon mass. The electric
and magnetic form factors, on the other hand, are suitable in extracting
them from the experiment: $ e^- N \rightarrow e^- N$ by Rosenbluch or polarization methods \cite{Rosenbluth}.
The four independent sets of form factors are related by
 \begin{eqnarray}
G_E^p(t) = F_1^p(t)+\tau^p F_2^p(t), \  \ \ G_M^p(t)=F_1^p(t)+F_2^p(t), \\ \nonumber
G_E^n(t)=F_1^n(t)+\tau^n F_2^n(t), \ \ \  G_M^n(t)=F_1^n(t)+F_2^n(t),     \nonumber
     \end{eqnarray}
      with $\tau^{p(n)}={t\over{4m_{p(n)}^2}}$,
     which can be interpreted as  Fourier transformations
      of the distribution of magnetism and charge in the Breit frame.
They satisfy the normalization conditions  $$ G_E^p(0)=1;\
G_M^p(0)=1+\mu_p;\ G^n_E(0)=0; G^n_M(0)=\mu_n;$$
 where $\mu_p$ and $\mu_n$ are the proton
and neutron 
 anomalous 
  magnetic moments, respectively.

 Since the GPD is not known a priori, one seeks for models of GPD
  based on general constraints on its analytic and asymptotic
 behavior. The calculated scattering amplitudes (cross sections) are than compared with the data to confirm, modify or reject the chosen form of the GPD.

  Commonly, the form $GPDs(x,\xi,t)$ is determined  through the
  exclusive deep inelastic processes  of type $\gamma^*p\rightarrow Vp,$ where $V$ stands for a  photon or  vector meson. However,
  such processes have a narrow region of momentum transfer and
  in most models the $t$-dependence of GPDs is taken in the factorization form
  with the Gaussian form of the $t$-dependence. Really, this form of $GPDs(x,\xi,t)$ can not be used
  to build the space structure of the hadrons,
   as for that one needs  to integrate over $t$ in a maximum wide region.

  The hadron
 form factors are related to the $GPDs(x,\xi,t)$
 by the 
 sum rules \cite{Ji97}
 \begin{eqnarray}
 F_1^q(t)=\int_{-1}^1 dx H^q(x,\xi=0,t),  \ \ \ \ F_1^q(t)=\int_{-1}^1 dx H^q(x,\xi=0,t).
 \end{eqnarray}

 The integration region can be reduced to positive values of
 $x,~0<x<1$ by the following combination of non-forward parton
 densities \cite{Rad1,GPRV}
  ${\cal H}^q(x,t)=H^q(x,0,t)+H^q(-x,0,t)$,
   ${\cal E}^q(x,t)=E^q(x,0,t)+E^q(-x,0,t)$,
    providing
 $F^q_1(t)=\int_0^1 dx {\cal H}^q(x,t) \label{01}$,
 $F^q_2(t)=\int_0^1 dx {\cal E}^q(x,t).\label{02}$

 The proton and neutron Dirac form factors are defined as
 \begin{eqnarray}
 F_1^p(t)=e_uF_1^u(t)+e_dF_1^d(t),  \ \ \ \ F_1^n(t)=e_uF_1^d(t)+e_dF_1^u(t),
 \end{eqnarray}
  where $e_u=2/3$ and
 $e_d=-1/3$ are the relevant quark electric charges.
 As a result, the $t$-dependence of the $GPDs(x,\xi=0,t)$
can be determined from the analysis of the nucleon form factors
for  which experimental data exist in a wide region of momentum transfer.
 It is a unique situation as it  unites the elastic and inelastic processes.

In the limit $t\rightarrow 0$ the functions $H^q(x,t)$ reduce to
 usual quark densities in the proton: $$ {\cal\
 H}^u(x,t=0)=u_v(x),\ \ \ {\cal H}^d(x,t=0)=d_v(x)$$ with the
 integrals $$\int_0^1 u_v(x)dx=2,\ \ \ \int_0^1 d_v(x)dx=1 $$
 normalized to the number of $u$ and $d$ valence quarks in the
 proton.

 However, 
 the "magnetic" densities ${\cal E}^q(x,t=0)\equiv {\cal E}^q(x)$
 cannot be directly expressed in
 terms of the known parton distributions; however, their
 normalization integrals $$\int_0^1{\cal E}^q (x)dx\equiv k_q $$
are constrained by the requirement that the values $F_2^p(t=0)$
and $F_2^n(t=0)$ are equal to the magnetic moments of the proton
and neutron, whence $k_u=2k_p+k_n\approx 1.673$ and
$k_u=k_p+2k_n\approx -2.033$ follow \cite{GPRV}.
This  helps us to obtain the corresponding PDFs by analysing the magnetic form factor
 of the proton and neutron \cite{GPD-ST-PRD09} 

 In \cite{GPD-ST-PRD09},
 the $t$-dependence of  GPDs  in the simplest form
\ba
{\cal{H}}^{q} (x,t) \  = q(x)_{nf} \   exp [  a_{+}  \
\frac{(1-x)^2}{x^{m} } \ t ];  \ \ \
{\cal{E}}^{q} (x,t) \  =  q(x)_{sf} \   exp [  a_{-}  \
\frac{(1-x)^2}{x^{m} } \ t ];
\label{GPD0}
\ea
 was researched.
   Complicated analysis  of  all available experimental data on the electromagnetic form factors of proton and neutron simultaneously
   allows one  to obtain the $t$-dependence of the $GPDs(x,\xi,t)$  \cite{GPD-PRD14}.    
   Different Mellin moments of GPDs give the form factors for  different reactions.
 If the first momentum of $GPDs(x,\xi,t)$ gives the electromagnetic form factors, then
    the integration of the second moment of GPDs over $x$ gives
      the momentum-transfer representation  of the so called gravitomagnetic form factors over $x,t$,
\ba
\int^{1}_{0} \ dx \ x \sum_{u,d}[{\cal{H}}(x,t) \pm {\cal{E}}(x,t)] = A_{h}(t) \pm B_{h}(t).
\ea
  which are connected with the energy-momentum tensor.

   Further development of the model requires  careful analysis of the momentum transfer form of
   the GPDs and a properly chosen form of the PDFs. In Ref. \cite{GPD-PRD14},  analysis of
   more than 24 different
   PDFs was performed. We slightly complicated the form of the GPDs in comparison with Eq.(\ref{GPD0}),
   but it is the simplest one compared to other works (for example, Ref. \cite{Diehl-Kroll}):
\ba
{\cal{H}}^{u} (x,t) = q(x)^{u}_{nf} \   e^{2 a_{H}  \ \frac{(1-x)^{2+\epsilon_{u}}}{(x_{0}+x)^{m}}  \ t };    \ \ \
{\cal{H}}^{d} (x,t) \  = q(x)^{d}_{nf} \   e^{2 a_{H} (1+\epsilon_{0}) (\frac{(1-x)^{2+\epsilon_{d}}}{(x_{0}+x)^{m}} ) \ t },
\label{t-GPDs-H}
\ea
\ba
{\cal{E}}^{u} (x,t) = q(x)^{u}_{fl} \   e^{2 a_{E}  \ \frac{(1-x)^{2+\epsilon_{u}}}{(x_{0}+x)^{m}}  \ t }, \ \ \
{\cal{E}}^{d} (x,t) = q(x)^{d}_{fl} \   e^{2 a_{E}(1+\epsilon_{0}) (\frac{(1-x)^{2+\epsilon_{d}}}{(x_{0}+x)^{m}} ) \ t },
\label{t-GPDs-E}
\ea
 where $q(x)^{u,d}_{fl}=q(x)^{u,d}_{nf} (1.-x)^{z_{1},z_{2}}$.

The ratio of $ \mu G_{E}/G_{M}$ for the proton and neutron cases is presented in Fig. 1.
      Our calculations reproduce the data obtained by the polarization method  quite well.
\begin{figure}
\includegraphics[width=.45\textwidth]{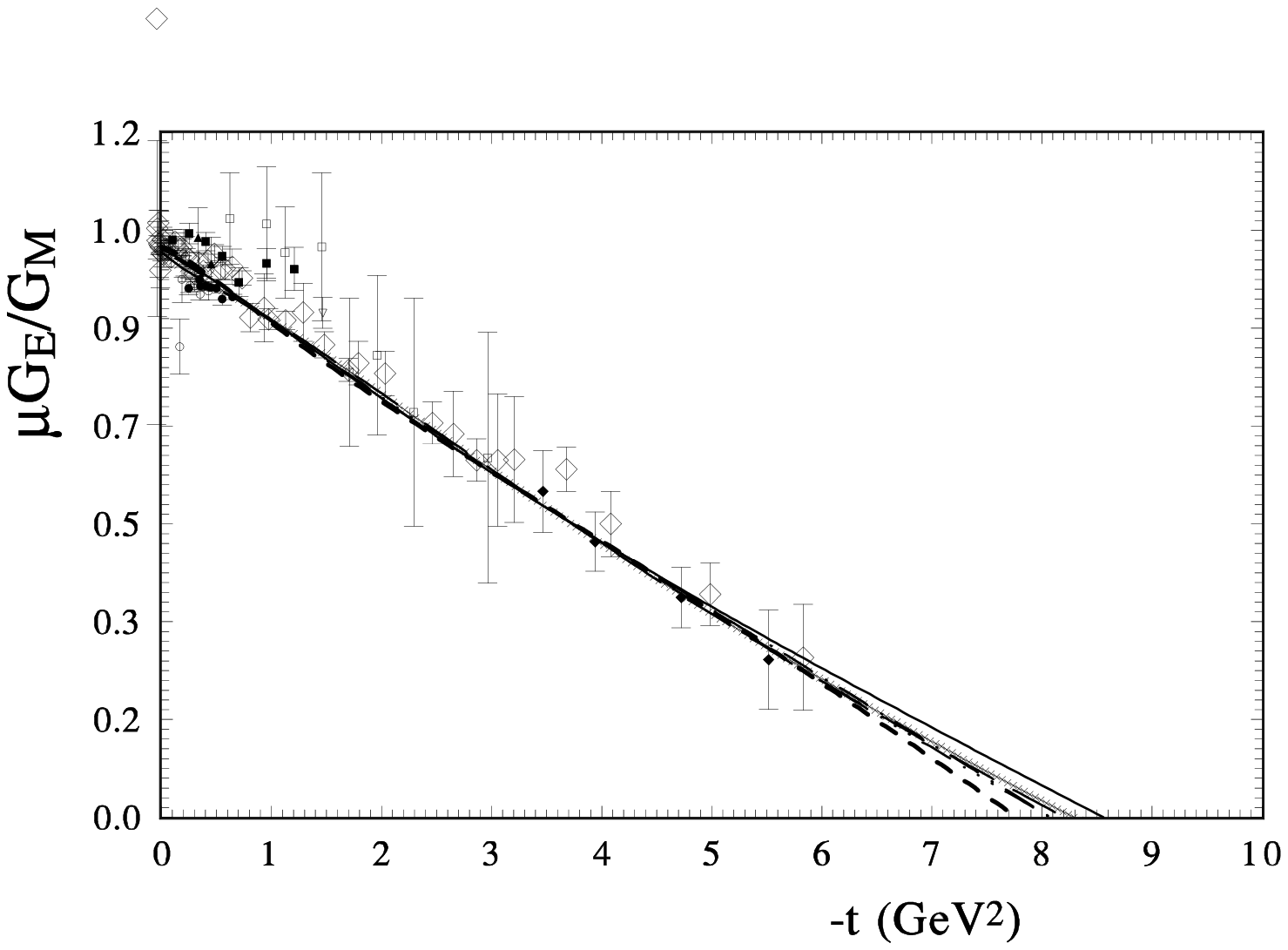} 
\includegraphics[width=.45\textwidth]{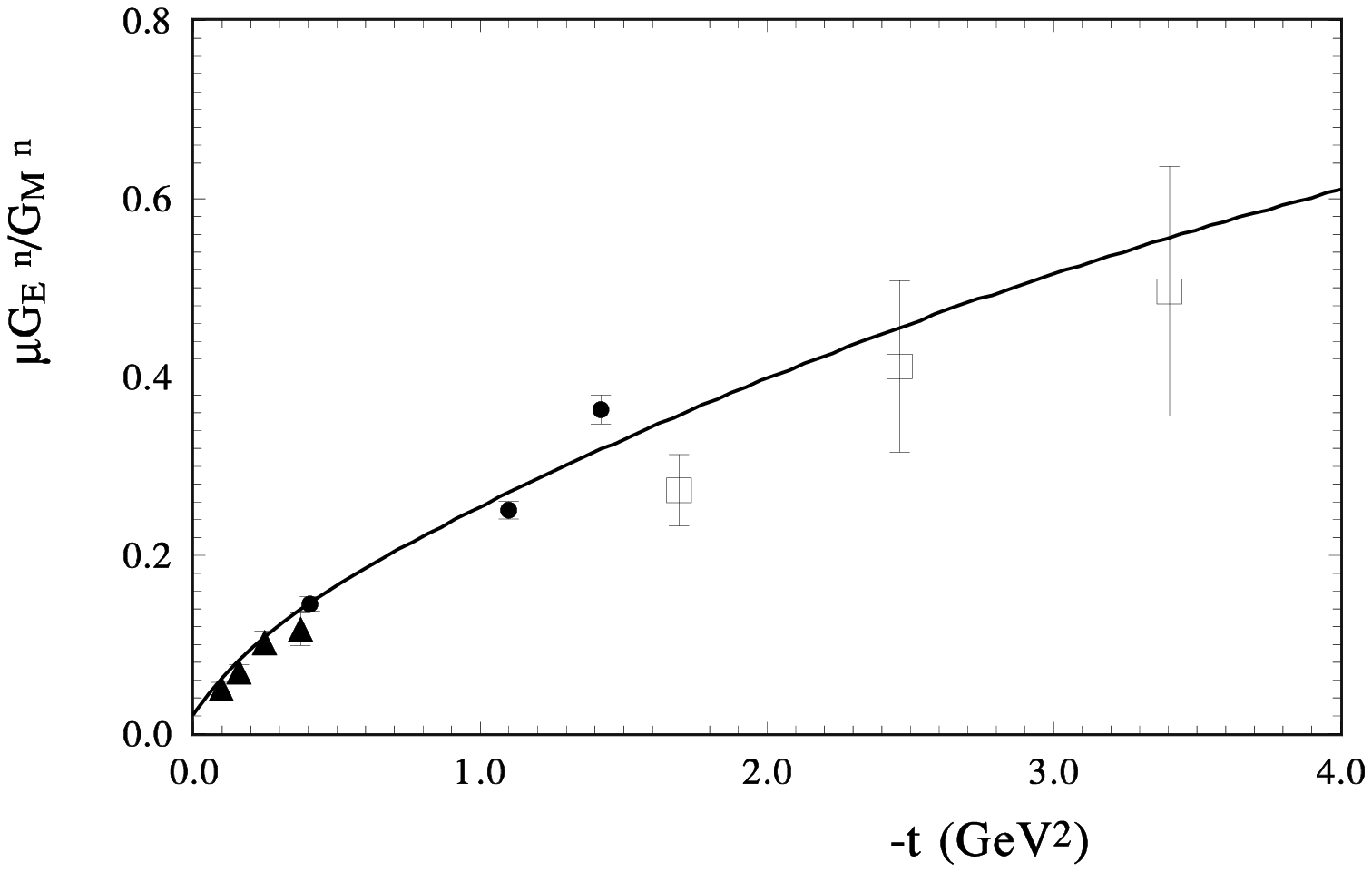} 
\vspace{1.cm}
\caption{The model description of the ratio of the electromagnetic form  factors
   for the proton $\mu_{p} G^{p}_{E}/G^{p}_{M}$ 
   with different forms of PDFs \cite{HEGS0} [left], 
      and
   for the neutron $\mu_{n} G^{n}_{E}/G^{n}_{M}$   [right],.
  }
\label{Fig3}
\end{figure}
 The hadron form factors   were calculated by
     using  numerical integration
 \ba
 F_{1}(t)= \int^{1}_{0}  dx
 [\frac{2}{3}q_{u}(x)e^{2 \alpha_{H} t (1.-x)^{2+\epsilon_{u}}/(x_{0}+x)^m}
   -\frac{1}{3} q_{d}(x)e^{ 2 \alpha_{H}  t (1.-x)^{2+\epsilon_{d}}/((x_{0}+x)^{m})} ]
\ea
   and then
    by fitting these integral results with the standard dipole form with some additional parameters
    for  $F_{1}(t)$,
 $  F_{1}(t)=  
   1/(1+q/a_{1}+q^{2}/a_{2}^2 +  q^3/a_{3}^3)^2 )$.  
  The matter form factor 
 \ba
 A(t) =  \int^{1}_{0} x \ dx
 [ q_{u}(x)e^{2 \alpha_{H} t (1.-x)^{2+\epsilon_{u}}/(x_{0}+x)^m}
   + q_{d}(x)e^{ 2 \alpha_{H}  t (1.-x)^{2+\epsilon_{d}}/((x_{0}+x)^{m})} ]
\ea
 is fitted   by the simple dipole form
$   A(t)  =  \frac{\Lambda^4}{(\Lambda^2 -t)^2 }$.
    The results of the integral calculations and the fitting procedure are shown in Fig.2.
        Our description is valid up to a large momentum transfer
        with the following parameters:
        $a_{1}=16.7,  \ a_{2}^{2}=0.78, \ a_{3}^{3}=12.5$ and $\Lambda^2=1.6$.
        These form factors will be used in our model of the proton-proton and proton-antiproton elastic scattering.

\begin{figure}
\includegraphics[width=.45\textwidth]{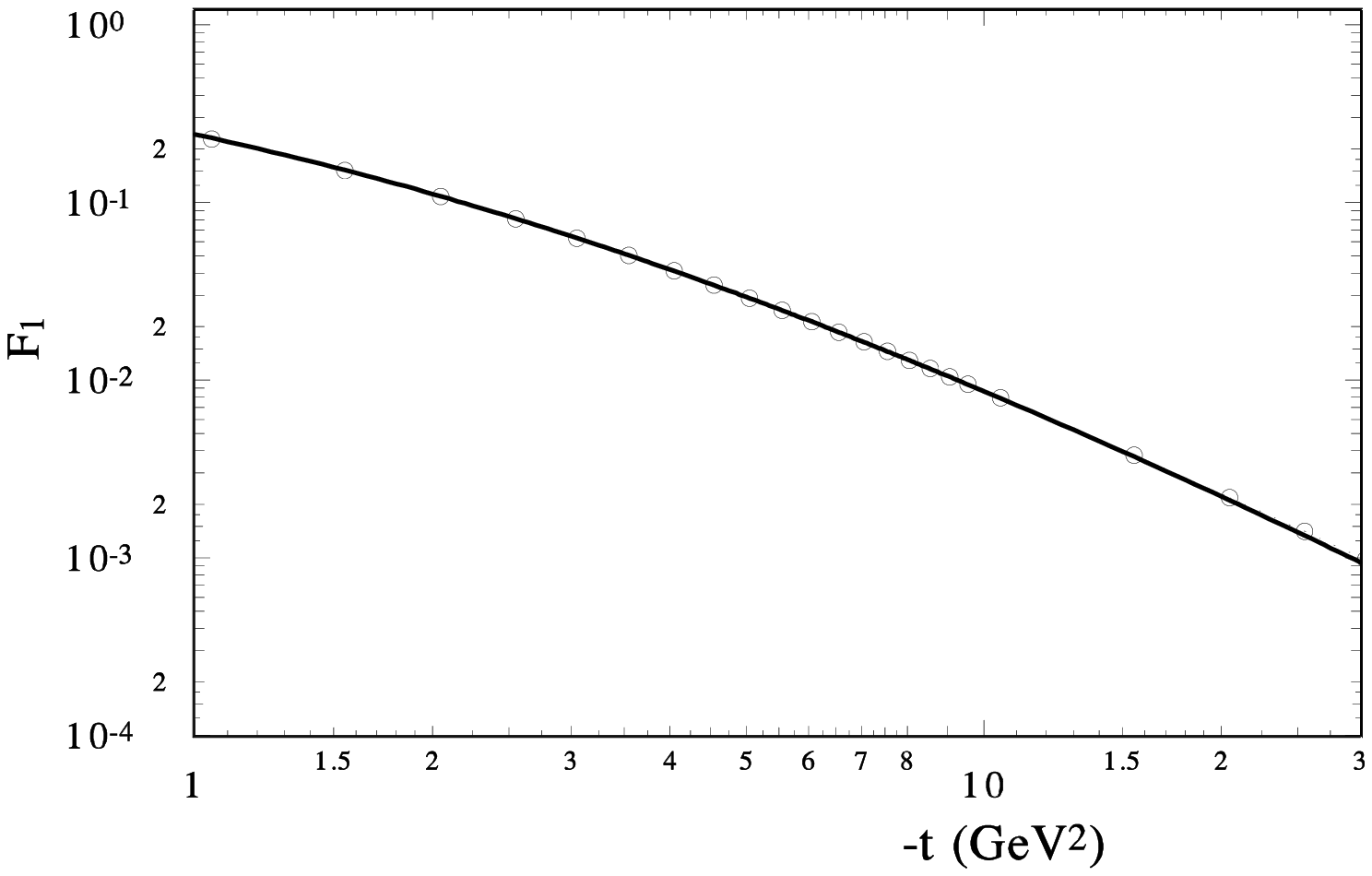} 
\includegraphics[width=.45\textwidth]{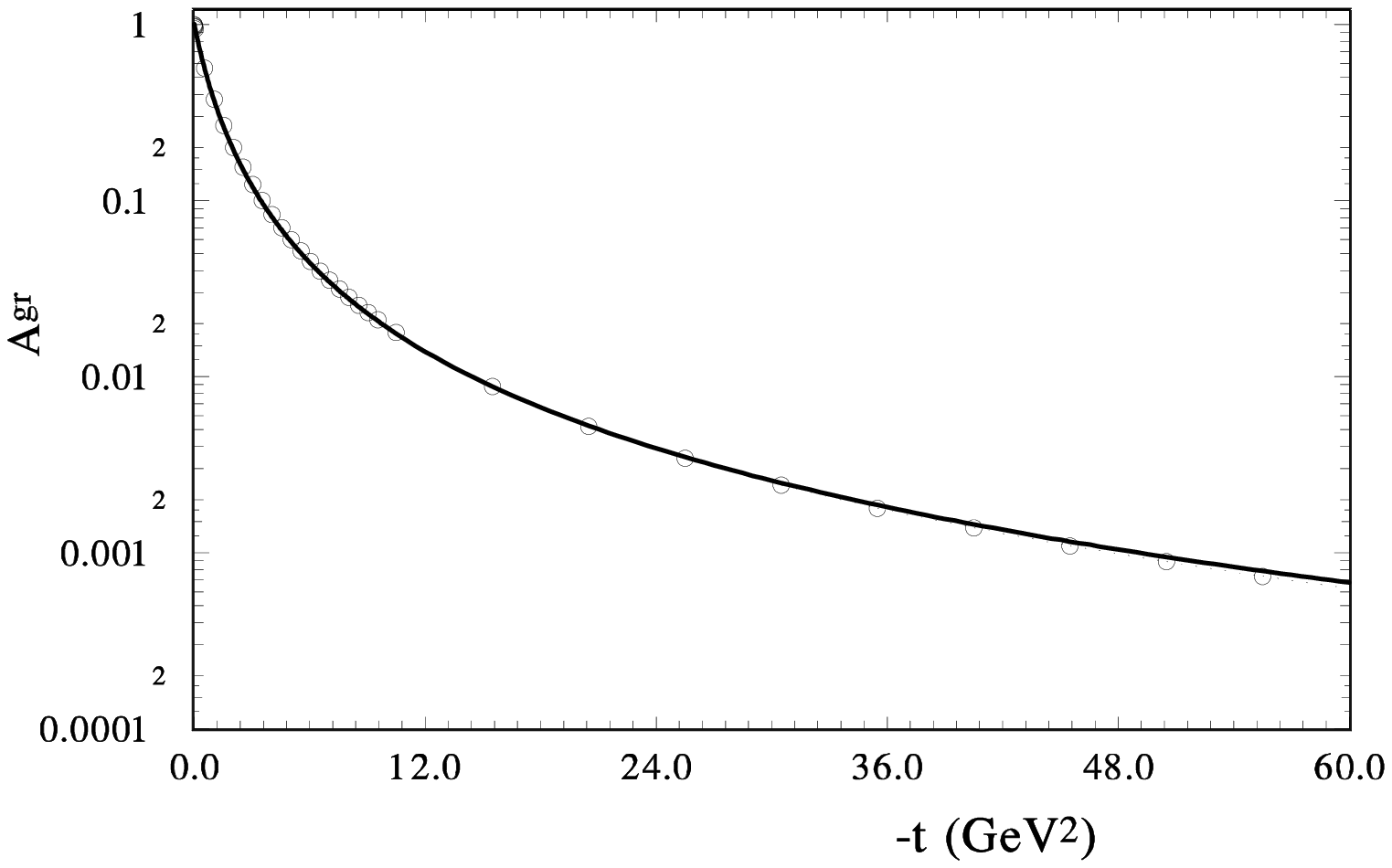} 
\vspace{1.cm}
\caption{ The fit of the form factors of the proton:
(a) [left], the electromagnetic form factor $G(t)$ 
 and
and [right] the matter form factor $A(t)$. 
 The circles are the moments of the GPDs (only every tenth point is shown).
  }
\label{Fig4}
\end{figure}

\section{Extension of the HEGS model with the spin-flip amplitude}

 In papers \cite{HEGS0,HEGS1}, the new High Energy Generelized Structure model was developed.
  The central moment of the model is that it 
  uses 
   two form factors  corresponding
   to charge and matter distributions calculated as the relevant moments of $GPDs(x,\xi=0,t)$.
   The basic Born spin-non-flip amplitudes were taken in the form
  \begin{eqnarray}
 F_{h}^{Born}(s,t) \ = h_1 \ G^{2}(t) \ F_{a}(s,t) \ (1+r_1/\hat{s}^{0.5}) 
    \    +  h_{2} \  A^{2}(t) \ F_{b}(s,t) \ (1+r_2/\hat{s}^{0.5}),
    \label{FB1}
\end{eqnarray}
  where $F_{a}(s,t)$ and $F_{b}(s,t)$  have the standard Regge form 
  \begin{eqnarray}
 F_{a}(s,t) \ = \hat{s}^{\epsilon_1} \ e^{B(s) \ t}, \ \ \
 F_{b}(s,t) \ = \hat{s}^{\epsilon_1} \ e^{B(s)/4 \ t}.
\label{FB-ab}
\end{eqnarray}
 The slope of the scattering amplitude has the  logarithmic dependence on energy,
 $   B(s) = \alpha^{\prime} \ ln(\hat{s})$,
 with  fixed $\alpha_{1}=0.24$ GeV$^{-2}$ and  $\Delta=0.11 $.
   Taking into account the Mandelstam region of the analyticity of  the scattering amplitude
   for the $2 \rightarrow 2 $ scattering process with identical mass
   $s+u+t = 4 m_{p}^2$,  one takes the normalized energy variable $s$ in a complex form $\hat{s}/s_{0}$ with
   $\hat{s} = s e^{i\pi}$, and $s_{0}=4 m_{p}^{2}$ where
     $m_{p}$ is the mass of the proton.
   In the present model, 
     a  small additional term is introduced into the slope, which reflects some possible small nonlinear
        properties of the intercept.
     As a result, the slope is taken in the form
  $ B(s,t) \ =  (\alpha_{1} + k q e^{-k q^2 Ln(\hat{s} \ t)} ) Ln(\hat{s}) $.
   This form leads to the standard form of the slope as $t \rightarrow 0$ and $t \rightarrow \infty$.
   Note that our additional term at large energies has a similar form as an additional term to the slope
   coming from the $\pi$ loop examined in Ref. \cite{Gribov-Sl} and recently in Ref. \cite{Khoze-Sl}.

      Then, as we intend to describe  sufficiently low energies,  possible Odderon contributions
      were taken into account: 
  \begin{eqnarray}
 F_{\rm odd}(s,t) \ =  \pm   \ h_{\rm odd} \ A^{2}(t) \ F_{b}(s,t),
\end{eqnarray}
 where $h_{\rm odd} = i h_{3} t/(1-r_{0}^{2} t) $.
  Just as we supposed in the previous variant of the HEGS model that  $F_{b}(s,t)$ 
    corresponds to the
   cross-even part of the three-gluon exchange, our Odderon contribution is also connected
   with the matter form factor $A(t)$.
   Our ansatz for the Odderon slightly differs from the cross-even part by some kinematic  
    function.
 The form of the Odderon working in  all $t$
    has the same behavior as the cross-even part at larger momentum transfer, of course,
   with different signs for proton-proton and proton-antiproton reactions.

 The final elastic  hadron scattering amplitude is obtained after unitarization of the  Born term.
    So, first, we have to calculate the eikonal phase
   \begin{eqnarray}
 \chi(s,b) \   = -\frac{1}{2 \pi}
   \ \int \ d^2 q \ e^{i \vec{b} \cdot \vec{q} } \  F^{\rm Born}_{h}\left(s,q^2\right)\,
 \label{chi}
 \end{eqnarray}
  and then obtain the final hadron scattering amplitude using eq.(9)
    \begin{eqnarray}
 F_{h}(s,t) = i s
    \ \int \ b \ J_{0}(b q)  \ \Gamma(s,b)   \ d b\,  \ \ \
 {\rm  with} \ \ \
  \Gamma(s,b)  = 1- \exp[\chi(s,b)].
 \label{Gamma}
\end{eqnarray}

Note that the parameters of the model are energy independent.
The energy dependence of the scattering amplitude is determined
only by the single intercept and the logarithmic dependence on $s$ of the slope.
   The analysis of the hard Pomeron contribution in the framework of the  model    \cite{NP-HP}
    shows that such a contribution is not felt. For the most part, the fitting procedure requires a negative  additional hard Pomeron contribution.
     We repeat the analysis of  \cite{NP-HP} in the present model
    and obtain practically the same results.
    Hence, we do not include the hard Pomeron in the model.

 Now we  do not know exactly, also from a theoretical
  viewpoint,  the dependence of  different parts of the
  scattering amplitude on $s$ and $t$. So, usually, we suppose
   that the imaginary and real parts of the spin-nonflip
  amplitude behave  exponentially  with the same slope, whereas the
  imaginary and real parts of  spin-flip amplitudes, without the
  kinematic factor $\sqrt{|t|}$, behave in the same manner with $t$ in the
  examined domain of transfer momenta.
  Moreover, one  mostly assume the energy independence of
  the ratio of the spin-flip to  spin-nonflip parts of the
  scattering amplitude.  All this is  our theoretical uncertainty.

  Let us take the main part of the spin-flip amplitude in the basic  form of the spin-non-flip amplitude.
  Hence, the born term of the spin-flip amplitude can be represented as
  \begin{eqnarray}
 F_{sf}^{Born}(s,t)= && h_{sf1} \ F_{1}^{2}(t) \ F_{sf-a}(s,t) \ (1+r_{sf1}/\hat{s}^{0.5})  \\ \nonumber
     + && h_{sf2} \  A^{2}(t) \ F_{sf-b}(s,t) 
     \pm h_{sf-odd} \  A^{2}(t)F_{sf-b}(s,t)\ (1+r_{sf2}/\hat{s}),  
    \label{FB}
\end{eqnarray}
  where $F_{sf-a}(s,t)$ and $F_{sf-b}(s,t)$  are the same as in the
   spin-non-flip amplitude
   but, according to the paper \cite{PS-Sl}, the slope
   of the amplitudes is essentially increasing. As a result, we take
   $F_{sf-a}(s,t) \ = \hat{s}^{\epsilon_1} \ e^{4 B(s) \ t}$,
   and $ F_{sf-b}(s,t) \ = \hat{s}^{\epsilon_1} \ e^{B(s)/2 \ t}$.
 It is to be noted  that most part of the available experimental data
 on the spin-correlation parameters exist only
  at sufficiently small energies.
   Hence, at  lower energies we need to take into account
  the energy-dependent parts of the spin-flip amplitudes.
  In this case,  some additional
  polarization data can be  included  in our examination.
  Then the spin-flip eikonal phase $\chi_{ls}(s,b)$ is calculated by the Fourier-Bessel transform, eq.(43),
   and then  the spin-flip amplitude in the momentum transfer representation
   is obtained by the standard eikonal representation for the spin-flip part, eq.(12).

  As in our previous  works \cite{Our-Kur,Nica20},  a small contribution from the energy-independent  part
  of the spin-flip amplitude in a form similar to that  proposed    in Ref. \cite{G-Kuraev2}  
  was added.
  \begin{eqnarray}
  F_{sf-t}(s,t) \ =  h_{sf} q^3 F_{1}^{2}(t) e^{-B_{sf} q^{2}}.
  \end{eqnarray}
  It has two additional free parameters. We take into account $ F_{sf-t}(s,t)$ and the full spin-flip amplitude is
  $ F_{sf}(s,t) = F_{sf-ab} + F_{sf-t}$.

   The model is very simple from the viewpoint of the number of fitting parameters and functions.
  There are no  artificial functions or any cuts which bound the separate
  parts of the amplitude by some region of momentum transfer.
We analyzed $3080$ experimental points
 in the energy region   $9.8$ GeV $\leq \sqrt{s} \leq 8 $ TeV
  and in the region of momentum transfer $0.000375 \leq |t| \leq 10 $ GeV$^2$
  for the differential cross sections
 and $125$ experimental points for the polarization parameter $A_N$ in the energy region
  $ 4.5 < \sqrt{s} < 30 $ GeV.
 The experimental data for the proton-proton and proton-antiproton elastic scattering are included
 in 87 separate sets of 30 experiments \cite{data-Sp,Land-Bron},
  including the recent data from the TOTEM Collaboration
 at $\sqrt{s}=8$ TeV  \cite{TOTEM-8nexp}.
  This gives us many experimental high-precision data points at a small momentum transfer, including
      the Coulomb-hadron interference region
      where the experimental errors are remarkably small.
      Hence, we can check  our model construction
      where the real part is determined only by the complex representation of $\hat{s}=s/s_{0} exp(-i \pi /2)$.
  We do not include the data on  the total cross sections
        $\sigma_{\rm tot}(s)$  and $\rho(s)$, as their values were obtained from the differential
        cross sections, especially in the Coulomb-hadron interference region.
         Including these data decreases $\chi^2$, but it  would be a double counting in our opinion.

 In the work, the fitting procedure  uses the modern version of the program
    "FUMILIM" \cite{Sitnik1,Sitnik2}" of the old program
    "FUMILY" \cite{fum83} which calculates the covariant matrix
    and gives the corresponding errors of the parameters and
    their correlation coefficients, and the errors of the final data.
    The analysis of the TOTEM data by three different statistical  methods, including the calculations through the correlation matrix  of the systematic errors,
     was made in \cite{GS-totan19}.

  As in the old version of the model, we  take into account only the statistical errors in the standard
      fitting procedure.
       The systematic errors are taken into account by the additional normalization coefficient
       which is the same for  every row of the experimental data.
       It essentially decreases the space of the possible form of the scattering amplitude.
        Of course, it is necessary to control the sizes of the normalization coefficients
         so that they do not introduce an additional energy dependence.
  Our analysis shows that the distribution of the coefficients has the
        correct statistical properties
      and does not lead to a visible additional energy dependence.
  As a result, we obtained a quantitative
    description of the experimental data ($\sum \chi^2/n_{dof} =1.3$).

%

\begin{figure}
  \includegraphics[width=.45\textwidth]{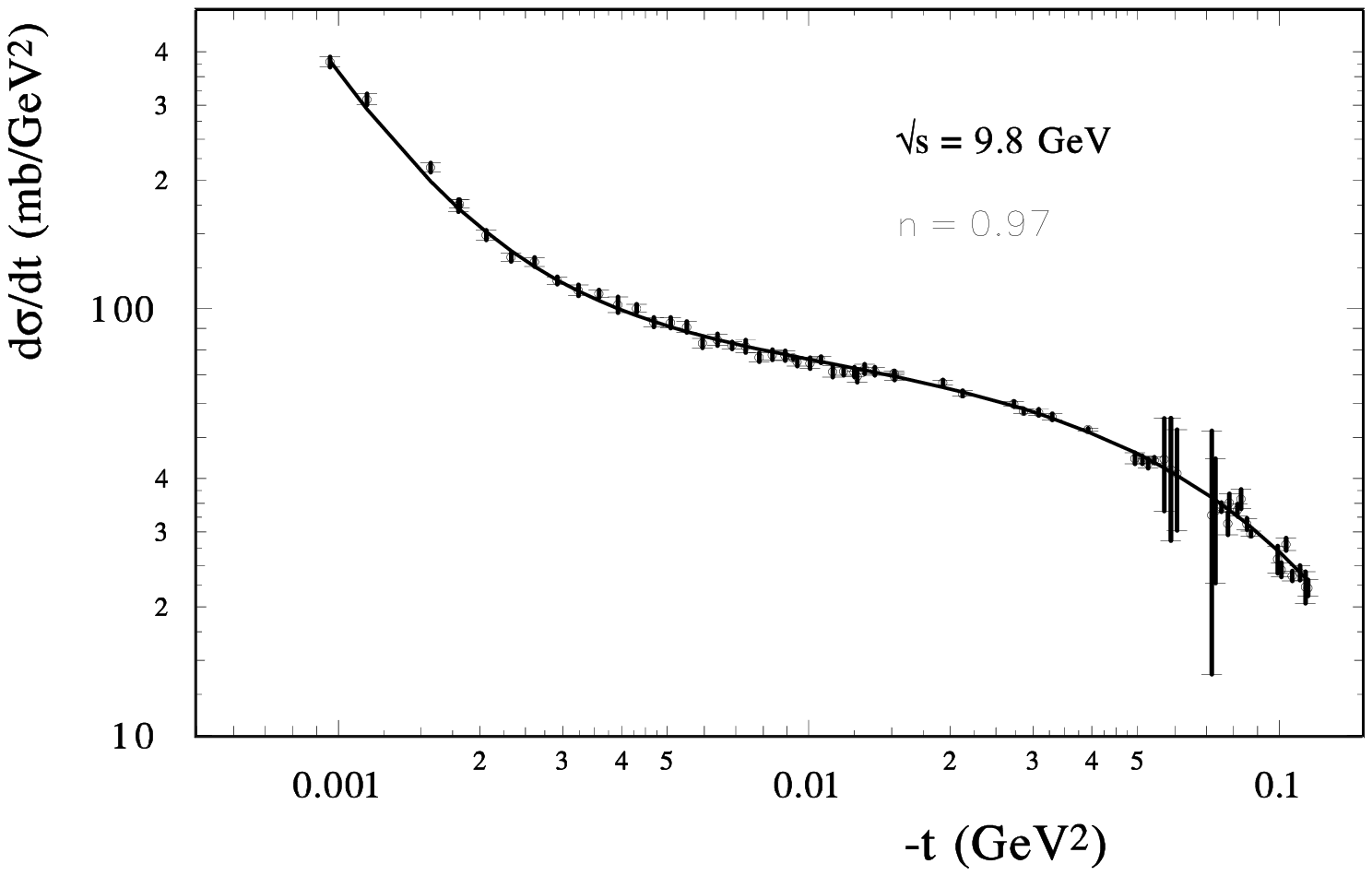}
      \includegraphics[width=.45\textwidth]{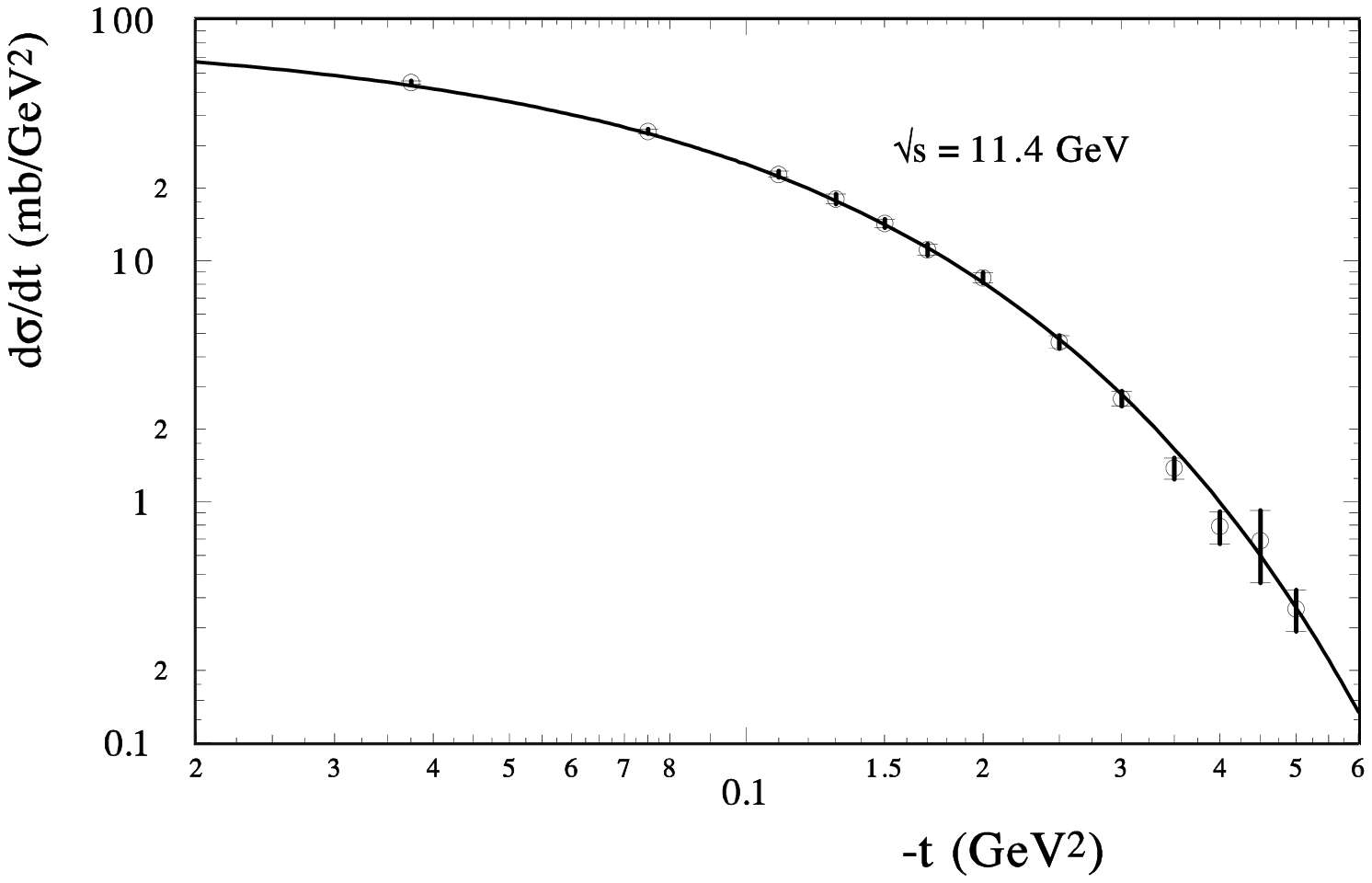}
\vspace{1.cm}
  \caption{
  $d\sigma/dt$ for $pp$ [left] at $\sqrt{s}=9.8$ GeV
    and $p\bar{p}$ [right] at $\sqrt{s}=11.4$ GeV (lines  - the model calculatios, points - the experimental data
    \cite{Whalley}).
   }
 \label{Fig5}
\end{figure}

  In the model, a good description of the CNI region of momentum transfer is obtained
  in a very wide energy region
   (approximately 3 orders of magnitude) with the same slope of the scattering amplitude.
 The differential cross sections 
 of the proton-proton and proton-antiproton elastic scattering
  at small  momentum transfer 
  are presented in  Fig. 3 at   $\sqrt{s}= 9.8 $ GeV   for $pp$ scattering,
   and $\sqrt{s}= 11 $ GeV for  $p\bar{p}$ elastic scattering.
   The model quantitatively reproduces the differential cross sections in the whole examined energy region
 in spite of the fact that the size of the slope  is essentially changing in this region
 [due to the standard Regge behavior $log(\hat{s}$]
  and the real part of the scattering amplitude has  different behavior for
   $pp$ and $p\bar{p}$.

\begin{figure*}
\begin{center}
\includegraphics[width=0.75\textwidth] {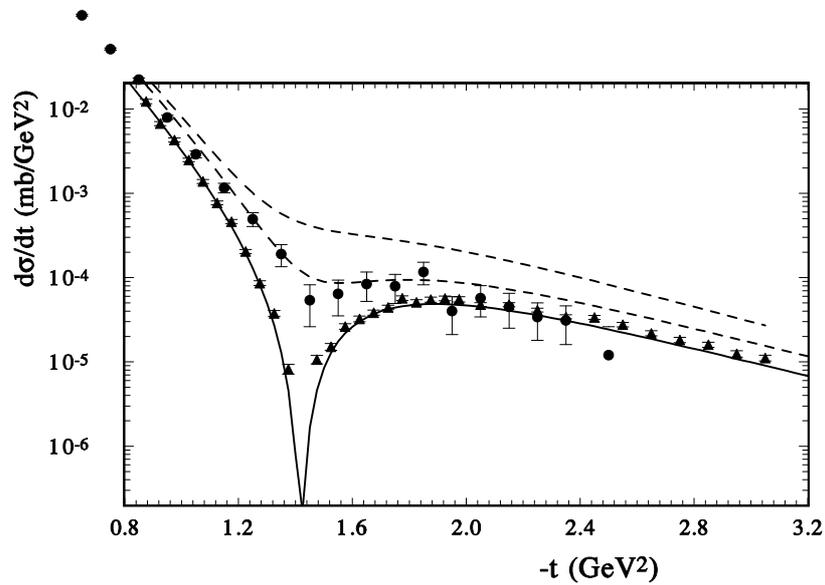}
\end{center}
\vspace{1.cm}
\caption{ The model calculation of the diffraction minimum in $d\sigma/dt$ of  $pp $ scattering
  at  $  \sqrt{s}=9.23, \ 13.76, \ 30.4  $~GeV; lines - the model calculations (shirt dash, long dash
  and solid; circles and triangles - the experimental data at 13.4 and 30.7 GeV
  \cite{Whalley}).
 }
\end{figure*}
%

 The form and the energy dependence of the diffraction minimum  are very sensitive
   to  different parts of the scattering amplitude. The change of the sign of the imaginary part
   of the scattering amplitude determines the position of the minimum and its movement
    with changing energy.
   The real part of the scattering amplitude determines the size of the dip.
   Hence, it depends    heavily on the odderon contribution.
   The spin-flip amplitude gives the contribution to the differential cross  sections additively.
   So the measurement of the form and energy dependence of the diffraction minimum
   with high precision is an important task for future experiments.
     In Fig.4, the description of the diffraction minimum in our model is shown for
     $\sqrt{s} = 9.23, \ 13.76$, and $\ 30.4 \ $GeV.
     The HEGS model reproduces  sufficiently well  the energy dependence and the form of the diffraction dip.
     In this energy region the diffraction minimum reaches the sharpest dip at  $\sqrt{s}=30 $~GeV.
     Note that at this energy the value of $\rho(s,t=0)$ also changes its sign in the proton-proton
     scattering.
     The $p\bar{p}$ cross sections in the model are obtained by the $s \rightarrow u$
       crossing without changing the model parameters.
      And  for the proton-antiproton scattering
       the same situation  with correlations between the sizes of  $\rho(s,t=0)$ and $\rho(s,t_{min})$
       takes  place at low energy (approximately  at $p_{L}= 100 $ GeV).
 Note that it gives a good description for the proton-proton and proton-antiproton elastic scattering
  or $\sqrt{s}=53 $~GeV and for $\sqrt{s}=62.1 $~GeV.
The diffraction minimum at $\sqrt{s}=7 $~TeV and $\sqrt{s}=13 $~TeV
 is reproduced sufficiently well too. 

   In the standard pictures, the spin-flip and double spin-flip amplitudes
    correspond to the spin-orbit $(LS)$ and spin-spin $(SS)$ coupling terms.
  The contribution to
  $A_N$ from the hadron double spin-flip amplitudes
   already at $p_L = 6 \ $GeV/c is of the second order
  compared to the contribution from the spin-flip amplitude.
   So with the usual high energy approximation for the helicity amplitudes
   at  small transfer momenta, we suppose that
   $\Phi_{1}=\Phi_{3}$ and we can neglect the contributions of the hadron parts
   of $\Phi_2-\Phi_4$.
   Note that if $\Phi_{1}, \Phi_3, \Phi_5$ have the same phases, their interference contribution
   to $A_N$ will be zero, though the size of the hadron spin-flip amplitude can be large.
   Hence, if this phase has  different $s$ and $t$ dependences, the contribution from the hadron
   spin-flip amplitude to $A_N$ can be zero at $s_i, \ t_i$ and non-zero at other $s_j, \ t_j$.

%
%
%
%
%
%
%
%
%
\begin{figure*}
\begin{center}
\includegraphics[width=0.45\textwidth] {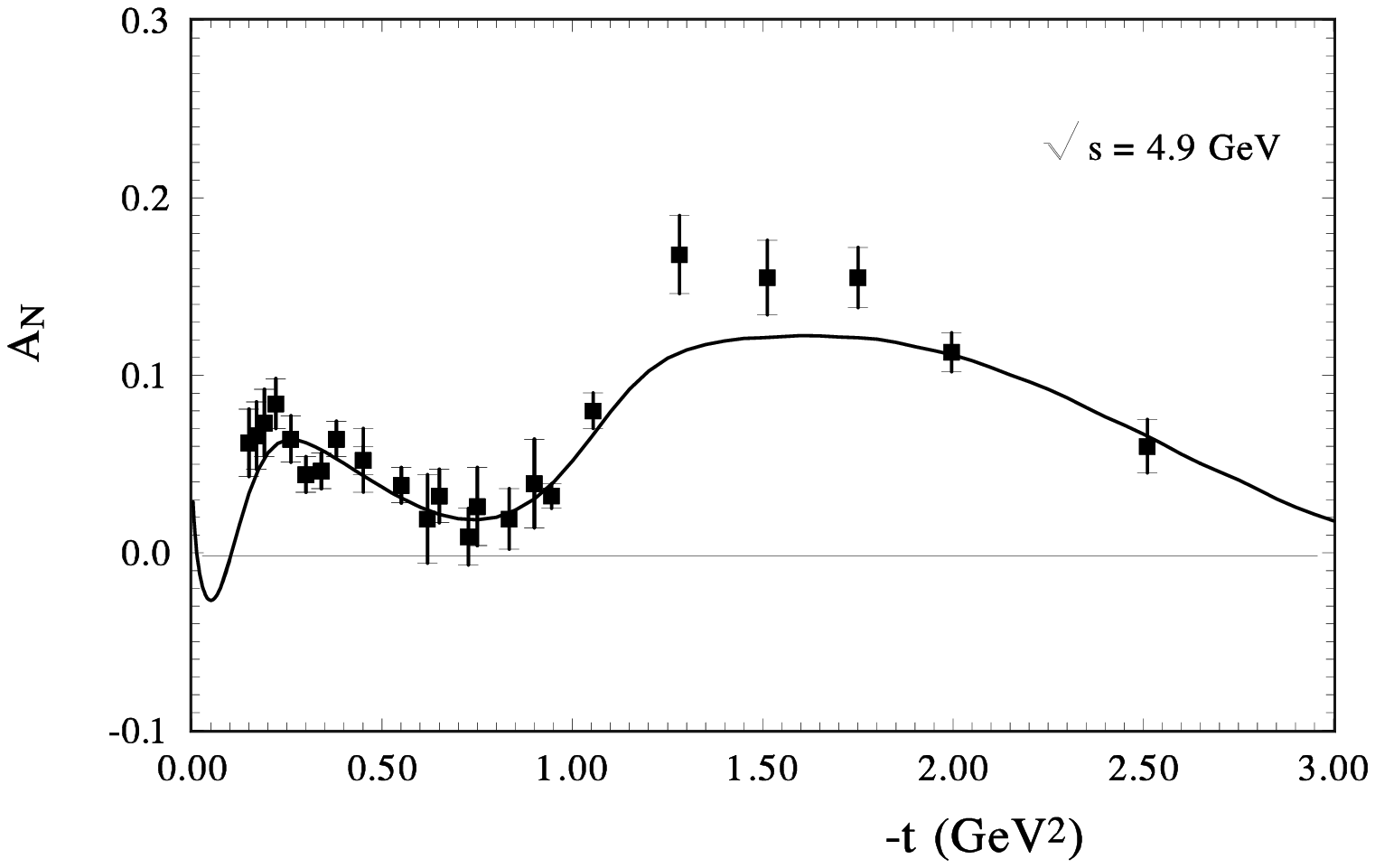}
\includegraphics[width=0.45\textwidth] {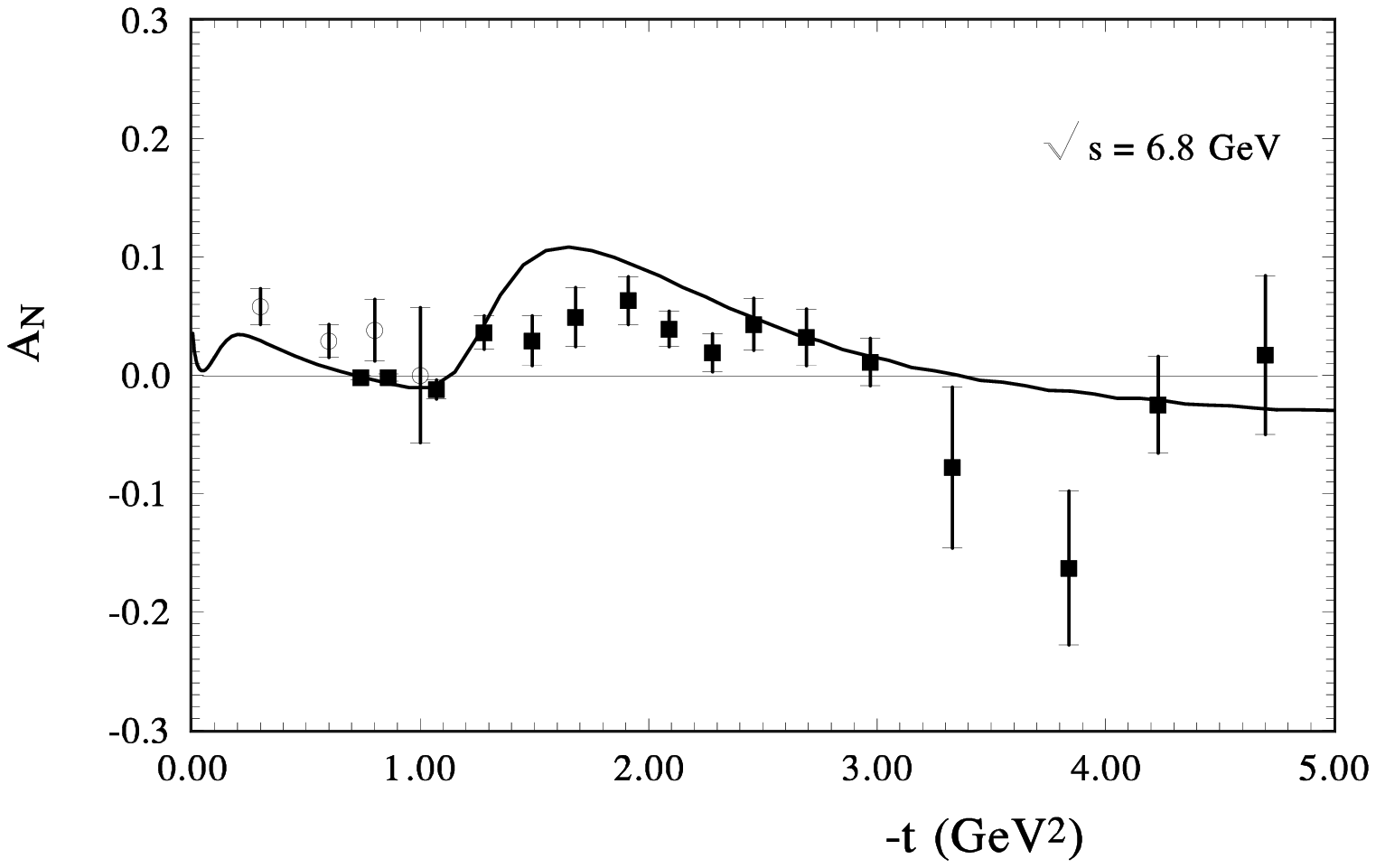}
\end{center}
\vspace{1.cm}
\caption{The analyzing power $A_N$ of pp - scattering
      calculated:
  a) at $\sqrt{s} = 4.9 \ $GeV  (the experimental data \cite{Pol4p9}),
   and
  b) at $\sqrt{s} = 6.8 \ $GeV    (points - the existing experimental data \cite{Pol6p8} ).
   }
\label{fig:10}       
\end{figure*}

\begin{figure*}
	\begin{center}
		\includegraphics[width=0.45\textwidth] {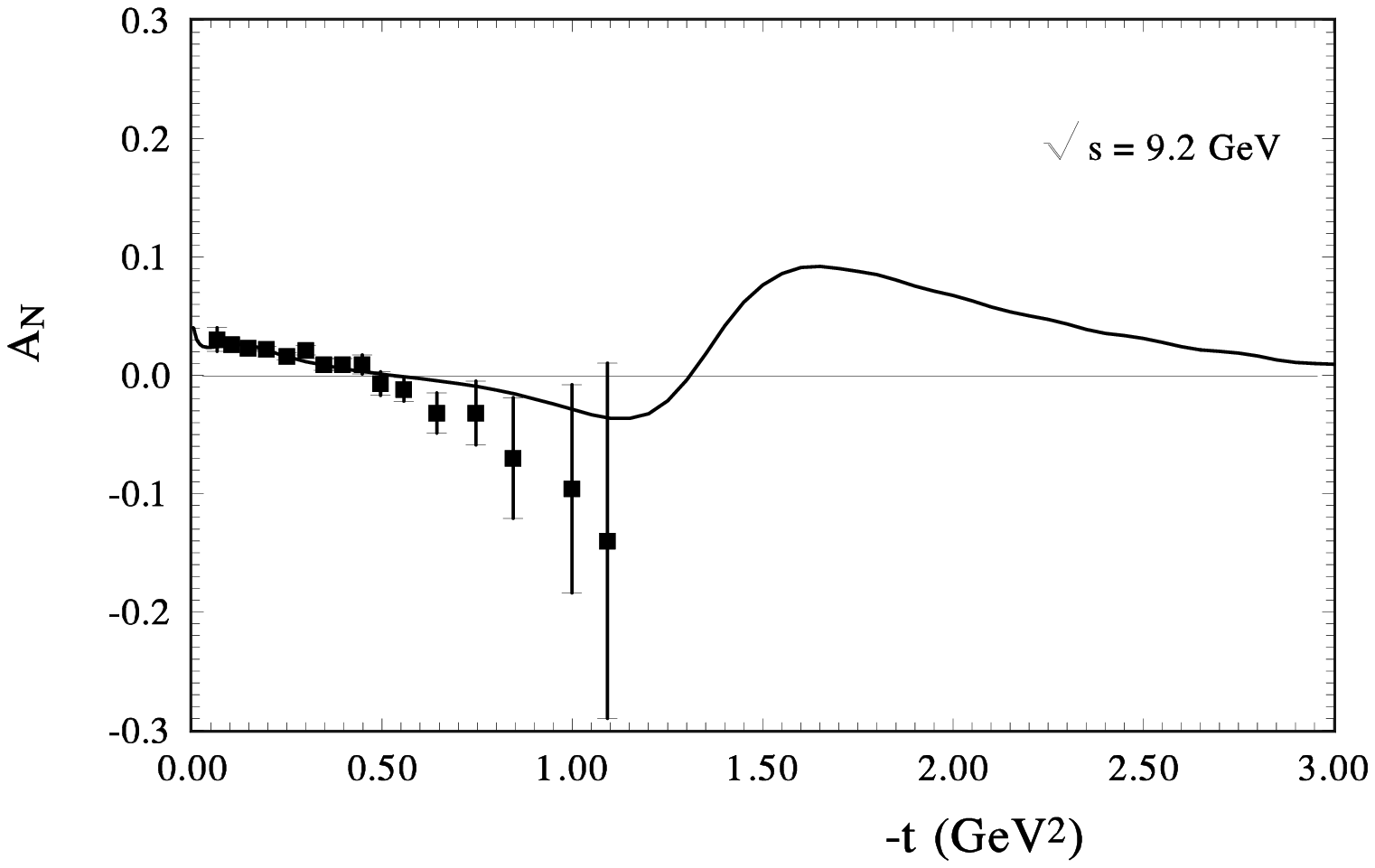}
		\includegraphics[width=0.45\textwidth] {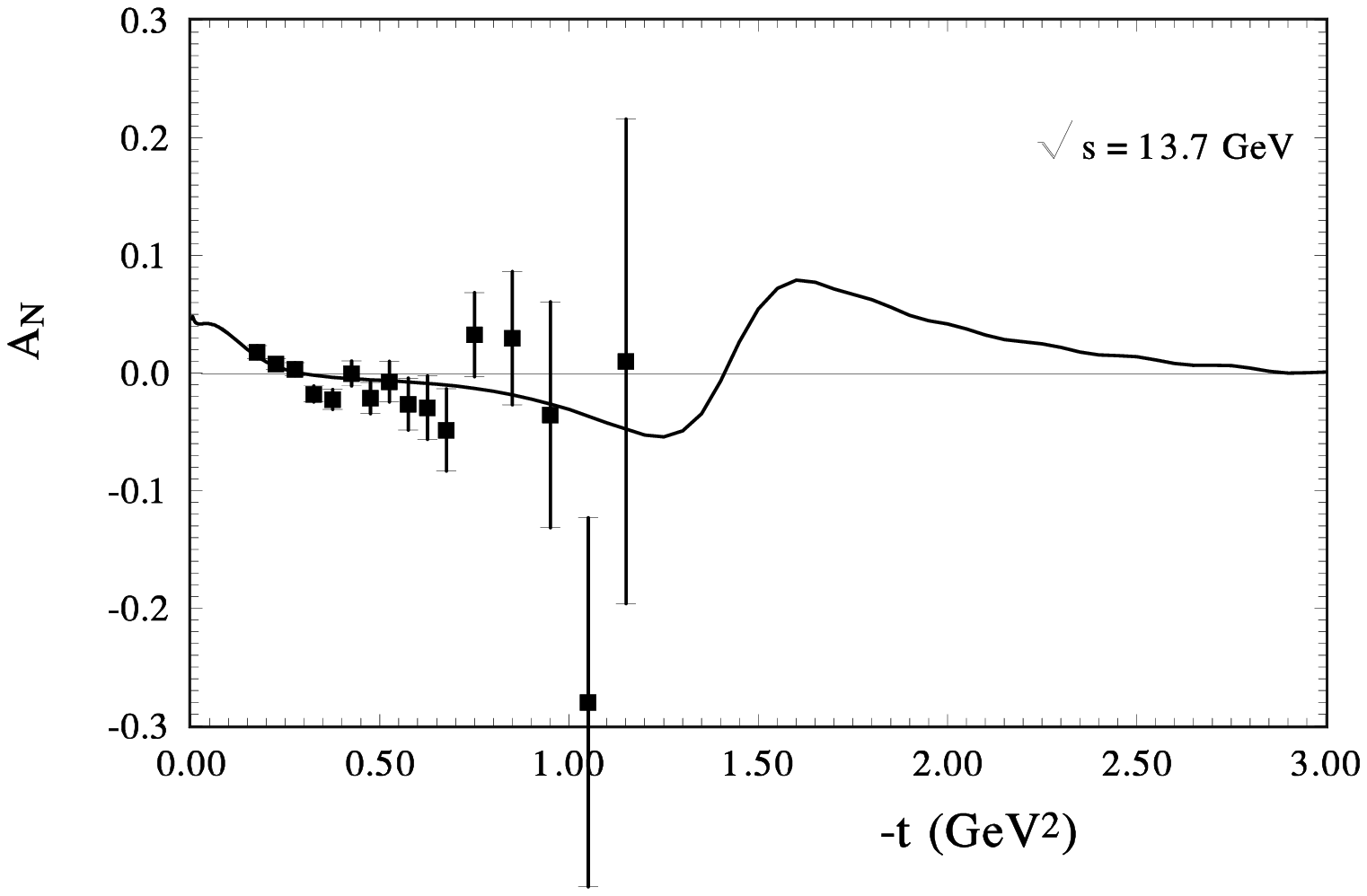}
	\end{center}
\vspace{1.cm}
	\caption{The analyzing power $A_N$ of pp - scattering
		calculated:
		a) at $\sqrt{s} = 9.2 \ $GeV,  (the experimental data \cite{Pol9p2}), and
		b)    at $\sqrt{s}= 13.7 \ $GeV (points - 
		the experimental data \cite{Pol23p4}). 
	}
	\label{fig:11}       
\end{figure*}

\begin{figure*}
\begin{center}
\includegraphics[width=0.45\textwidth] {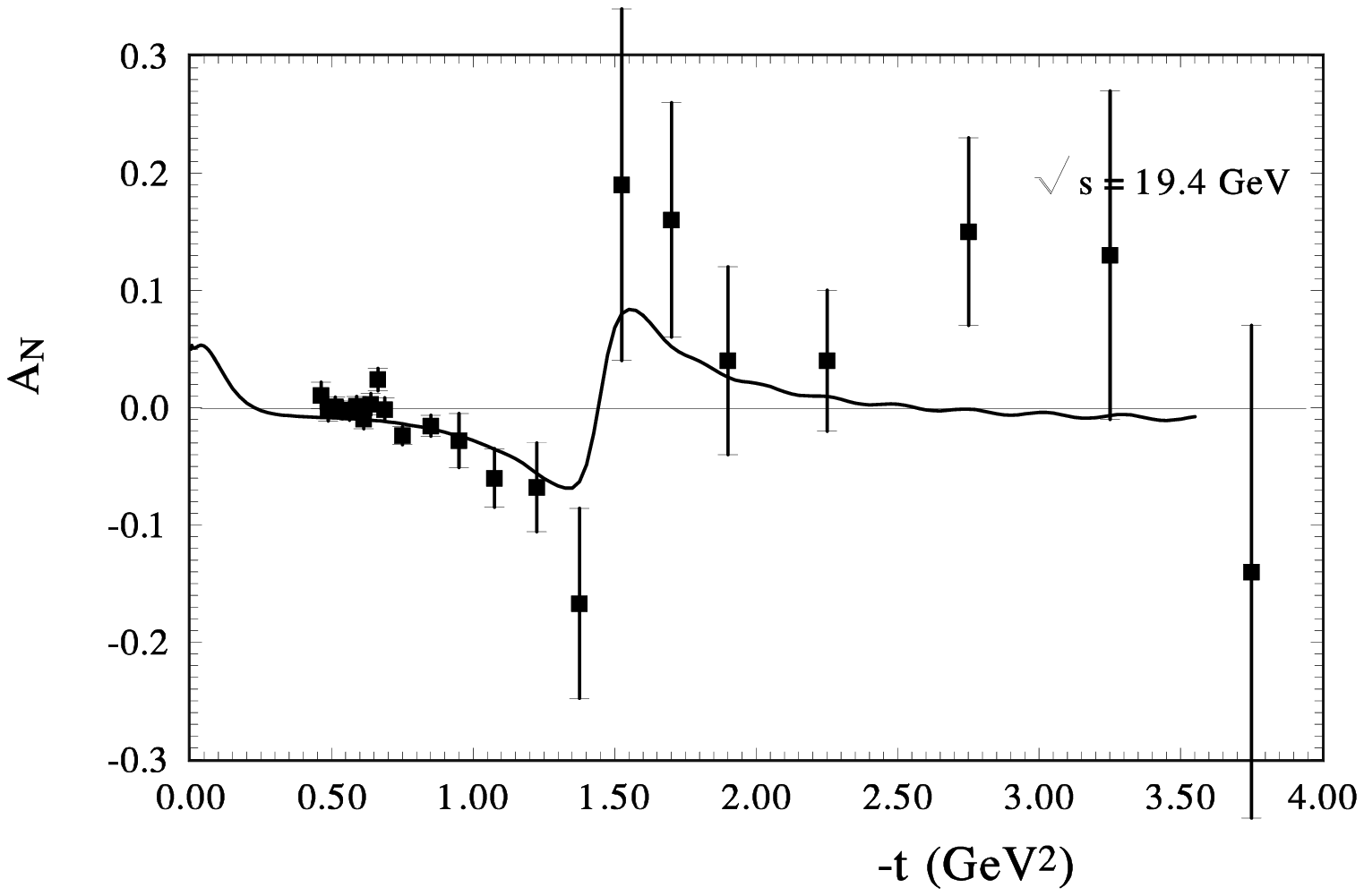}
\includegraphics[width=0.45\textwidth] {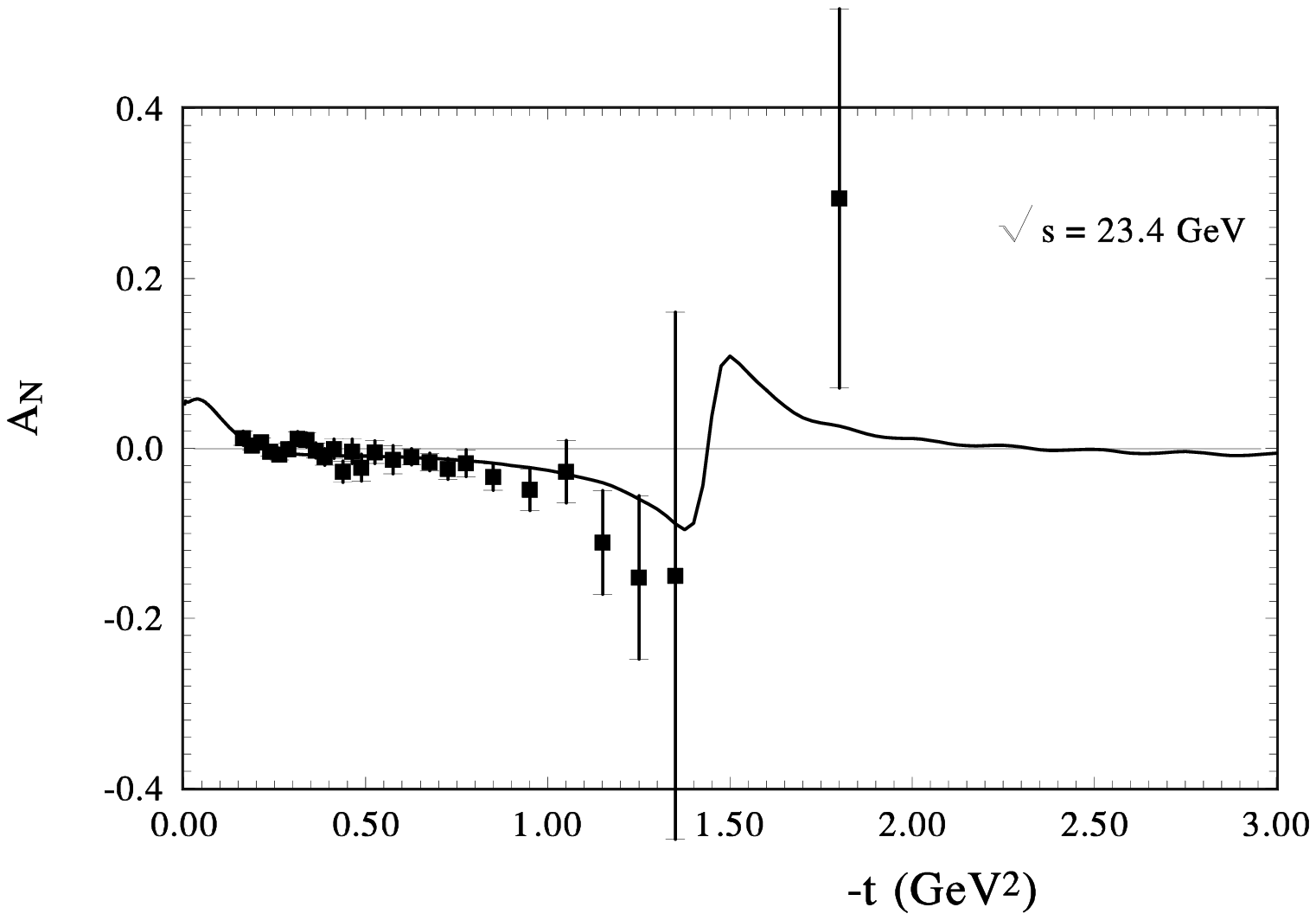}
\end{center}
\vspace{1.cm}
\caption{The analyzing power $A_N$ of pp - scattering
      calculated:
   a) at $\sqrt{s} = 19.4 \ $GeV  (the experimental data \cite{Pol19p4}),
   and
   b)    at $\sqrt{s}= 23.4 \ $GeV
       (points - the existing experimental data \cite{Pol23p4})
          }
\label{fig:2}       
\end{figure*}

\begin{figure}
%
\begin{flushright} 
\includegraphics[width=.45\textwidth]{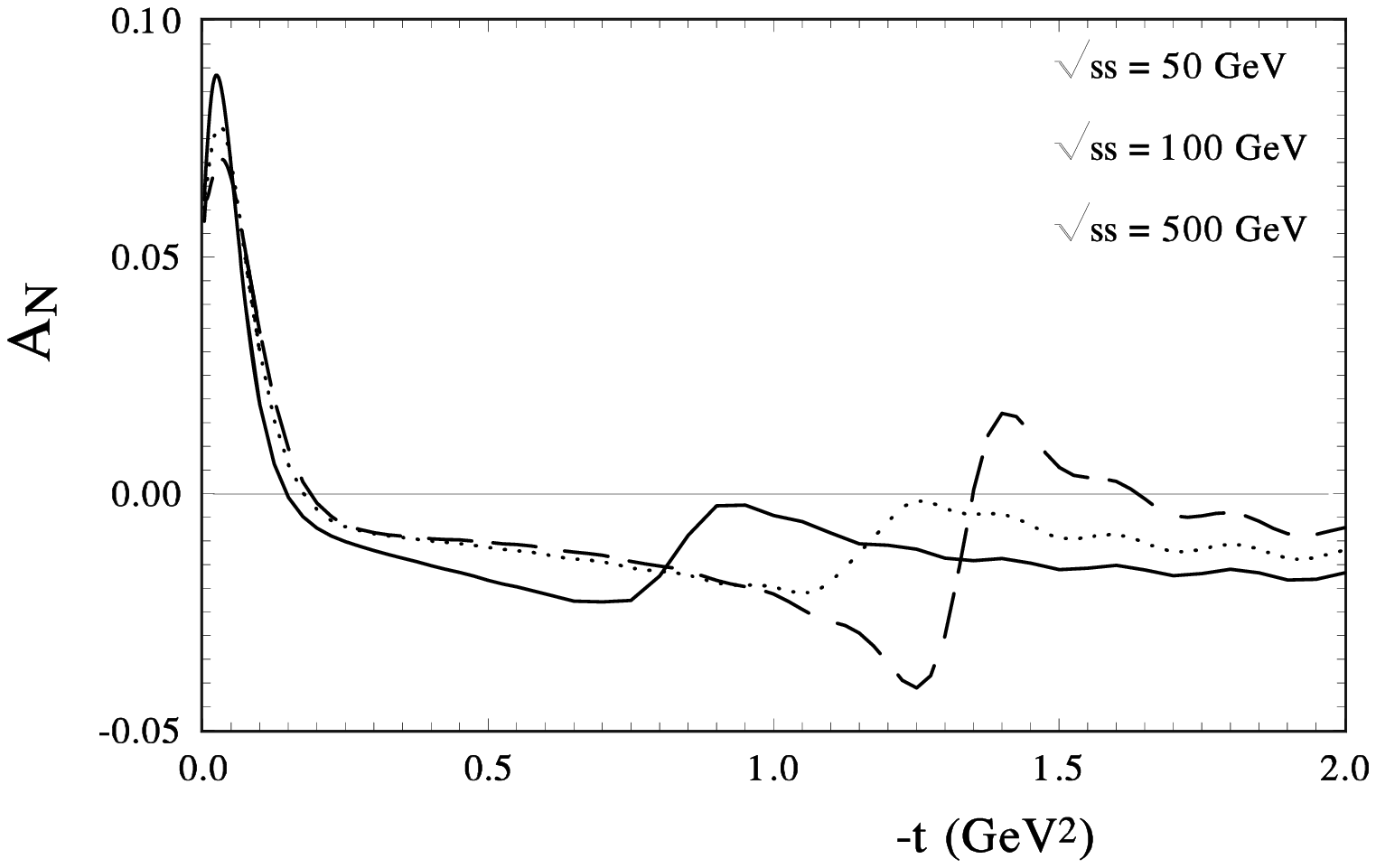}
\includegraphics[width=0.45\textwidth]{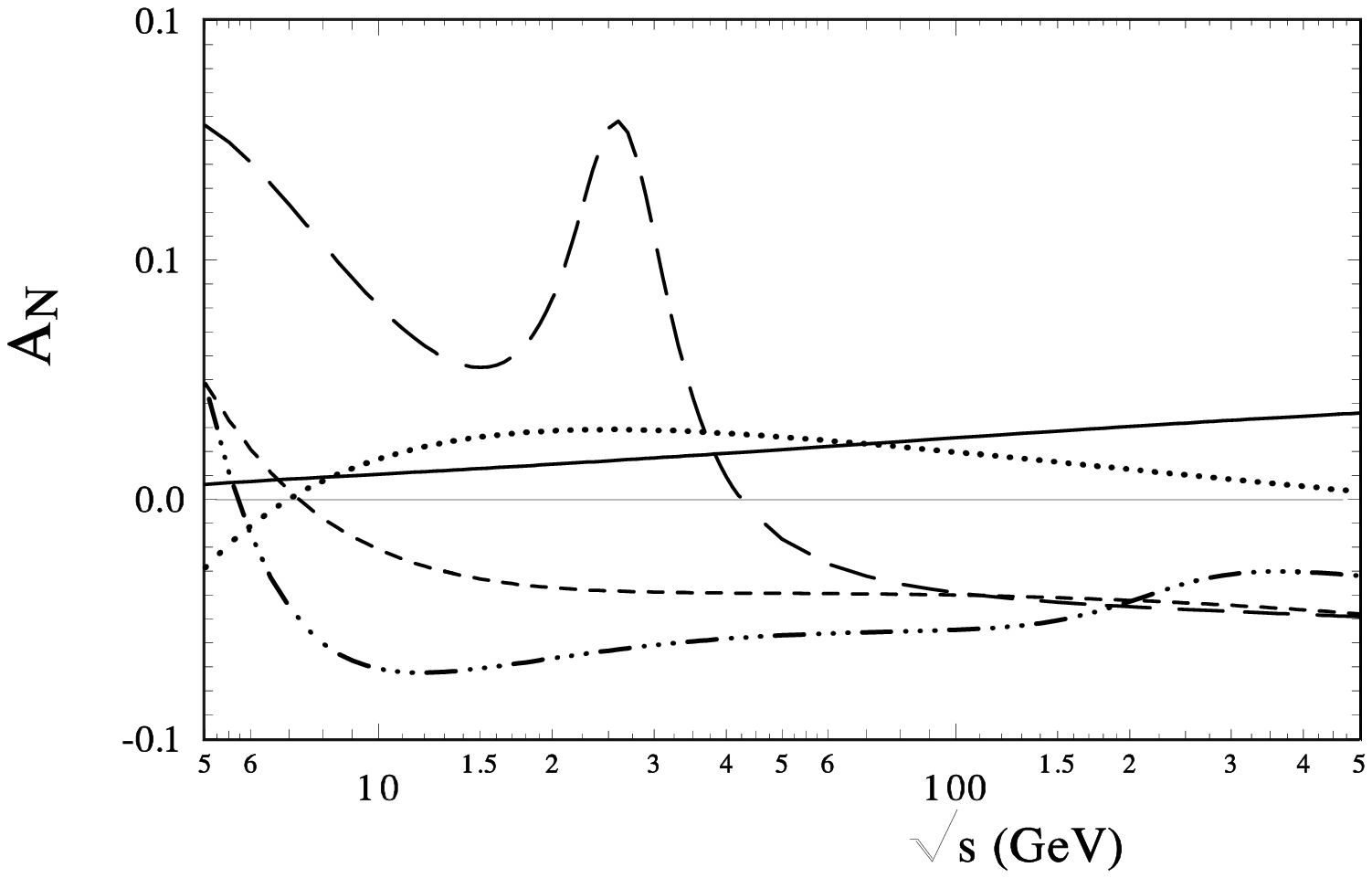}
\end{flushright}
\vspace{1.cm}
\caption{[left]  The calculated size of the spin correlation parameter $A_{N}(s,t)$
 at high energies $\sqrt{s}= 50, \ 100, \ 500 \ $GeV.
 [right] the $s$-dependence of $A_{N}(s,t)$ at different fixed $t_{i} = 0.001, \ 0.1, \ 0.4, 1.0, 1.5 \ $ GeV
 (solid, dots, short dash, dot-dot-dash and long-dash lines, respectively).
  }
\label{Fig_3}
\end{figure}


\begin{figure}
%
\begin{flushright} 
\includegraphics[width=.45\textwidth]{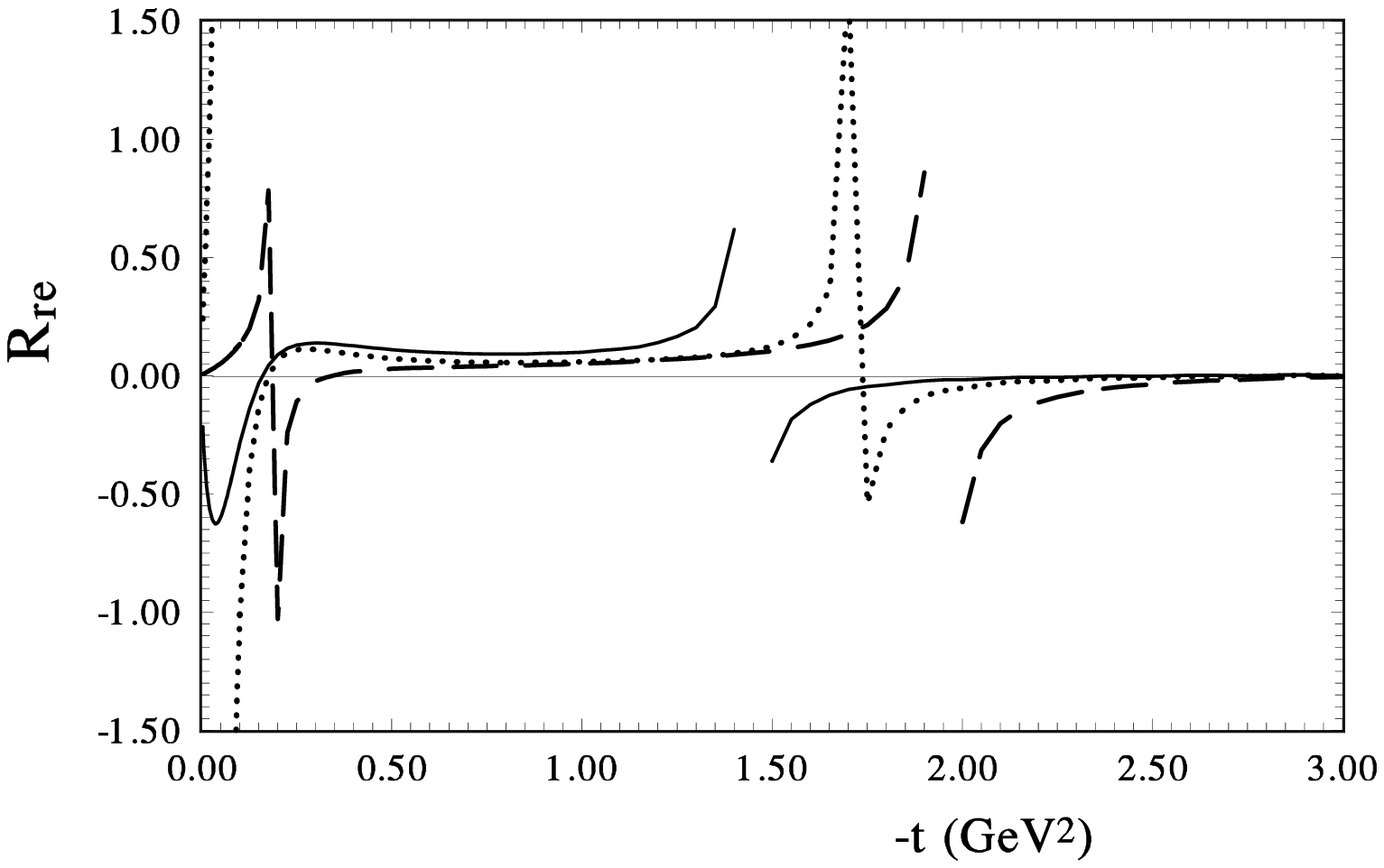}
\includegraphics[width=0.45\textwidth]{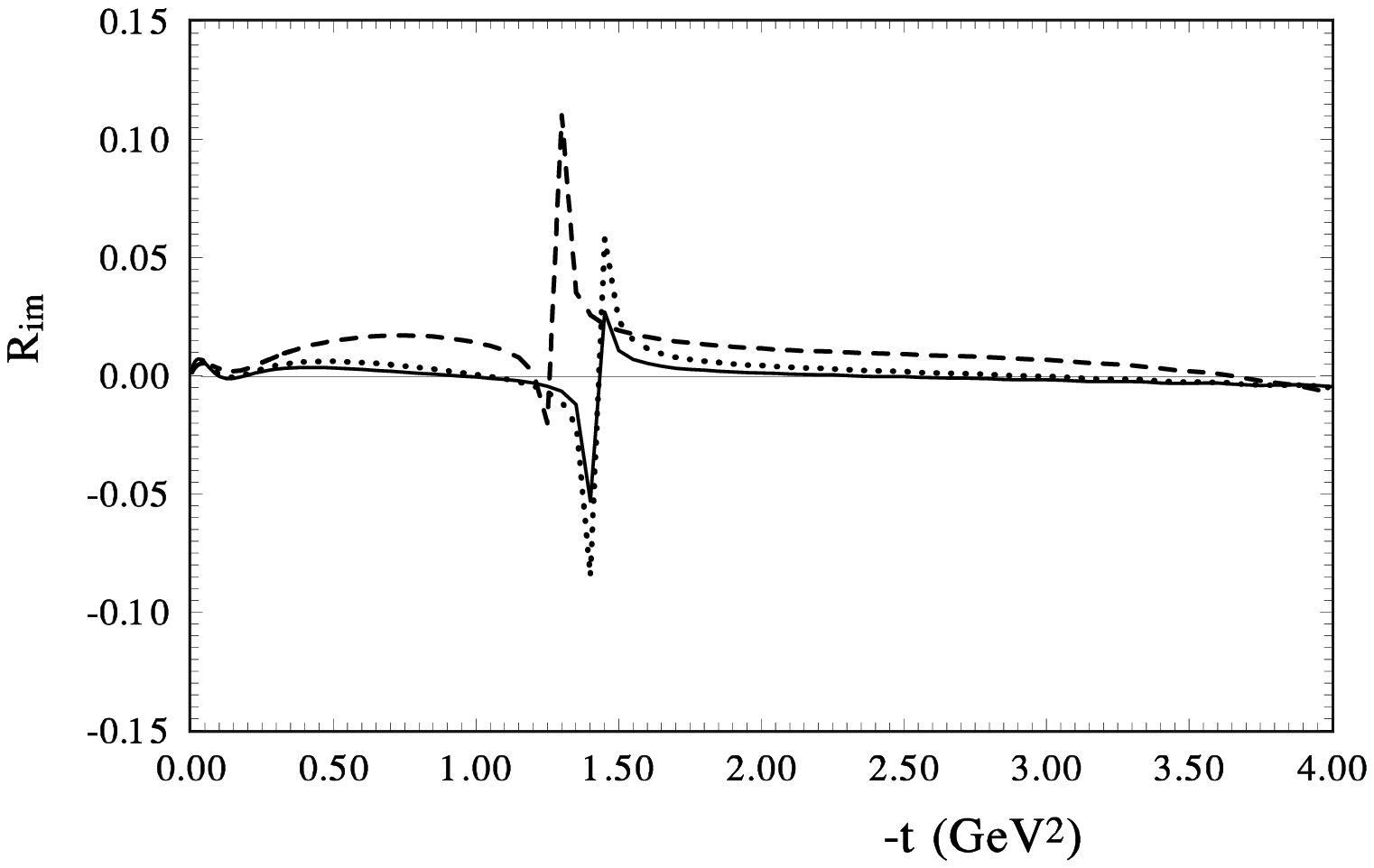}
\end{flushright}
\vspace{1.cm}
\caption{ [left] The ratio of the imaginary parts of the spin-flip amplitude and
   spin-non-flip amplitude; [right] the ratio of the real parts of the spin-flip amplitude and
   spin-non-flip amplitudes (lines - dashed, dots, and solid correspond to $\sqrt{s} = 9.23, \ 19.4, \ 30.7 \ $GeV).
  }
\label{Fig_4}
\end{figure}

Our calculation for $A_{N} (t)$ is shown in Fig. 5 a,b. at $\sqrt{s}=4.9 $~GeV and $\sqrt{s}=6.8 $~GeV.
 For our high energy model it is a very small energy. However,
  the description of the existing data is sufficiently good. At these energies, the diffraction
  minimum practically is overfull by the real part of the spin-non-flip amplitude and
  the contribution of the spin-flip-amplitude; however, the $t$-dependence of
the  analysing power is very well reproduced in this region of the momentum transfer.
Note that the magnitude and the energy dependence of this parameter depend on
the energy behavior of  the zeros
of the imaginary-part of the spin-flip amplitude
and the real-part of the spin-nonflip amplitude.
Figure 6 shows  $A_{N} (t)$ at $\sqrt{s}=9.2 $~GeV and $\sqrt{s}=13.7 $~GeV.
 At these energies the diffraction minimum deepens and its form affects  the form of $A_{N} (t)$.
 At last,  $A_{N} (t)$ is shown at large energies  $\sqrt{s}=19.4 $~GeV and $\sqrt{s}=23.4 $~GeV
 in Fig.7.
 The diffraction dip in the differential cross section has a sharp form and it affects  the
 sharp form of $A_{N} (t)$.
The maximum negative values of $A_N$  coincide closely with
the diffraction minimum.   
  We found that the contribution
of the spin-flip to the differential cross sections is much less
than the contribution of the spin-nonflip amplitude in the examined
region of momentum transfers from these figures; $A_N$ is determined
in the domain of the diffraction dip by the ratio
\begin{eqnarray}
                 A_N \sim Im f_{- } / Re f_{+}. \label{ir}
\end{eqnarray}
The size of the analyzing power changes from $-45\%$ to $-50\%$
at $\sqrt{s}=50 \ \gev$ up to $-25\%$ at $\sqrt{s}= 500 \ \gev$.
These numbers give the magnitude of the ratio Eq.(\ref{ir}) that does
not strongly depend on the phase between the spin-flip and
spin-nonflip amplitudes.  This picture implies that the diffraction minimum
is   mostly filled by the real-part of the spin-nonflip amplitude
and that the imaginary-part of the spin-flip amplitude
increases in this domain as well.

We observe that the dips are
different in speed of displacements with energy from Fig. 4.  
In Fig. 8,  one sees that at larger momentum transfers,
$|t| \sim 2$ to $3 \ \gvs$, the analyzing power depends on energy
very weakly.
   The spin-flip amplitude gives the contribution to the differential cross  sections additively.
   So the measurement of the form and energy dependence of the diffraction minimum
   with high precision is an important task for future experiments.

   In Fig.8, the predictions of the HEGS model are presented  for $A_{N} (t)$  up to
     high energies $\sqrt{s}=500 $~GeV. 
     It can be seen
     that at such huge energy the size of $A_{N} (t)$ does not come to zero and
     can be measured in new   LHC experiments with a fixed target.

    Now let us examine the ratio of the real and imaginary
   parts of the spin-flip
    phenomenological and model amplitudes to their
   imaginary parts of the hadron spin-non-flip amplitudes
    (see Figs. 9 (a,b) and 11 (a,b)).
    It is clear that this ratio can not be regarded as a constant.
   Moreover, this ratio has a
  very strong energy dependence.

   Neglecting  the $ \Phi_{2}(s,t)- \Phi_{4}(s,t)$  contribution, the spin correlation parameter $A_{N}(s,t)$
   can be written taking into account the phases of  separate amplitudes
  \begin{eqnarray}
  A_{N}(s,t)  \frac{d \sigma}{dt} = -\frac{4 \pi}{s^2}
    [|F_{nf}(s,t)| \ |F_{sf}(s,t)| Sin( \theta_{nf}(s,t)-\theta_{sf}(s,t))].
\end{eqnarray} 
    where  $\theta_{nf}(s,t), \theta_{sf}(s,t)$ are the phases of the spin non-flip and spin-flip amplitudes.
  It is clearly seen that despite the large spin-flip amplitude the analyzing power can be near zero
  if the difference of the phases is zero in some region of momentum transfer.
  The experimental data at some point of the momentum transfer show the energy independence of
  the size of the spin correlation parameter $A_{N}(s,t)$. 
  Hence, the small value of the $A_{N}(s,t)$ at some $t$ (for example, very small $t$)
  does not serve as a proof that it will be small in other regions of momentum transfer.

%

  Let us compare the  spin-flip amplitudes  
  and the spin-nonflip amplitudes   in the impact parameter representation
   at  $\sqrt{s}=30$ GeV.
 The results are present in Fig.10.
 It can be seen that the first has more peripheral behavior.
\begin{figure}
\begin{center}
\includegraphics[width=.5\textwidth]{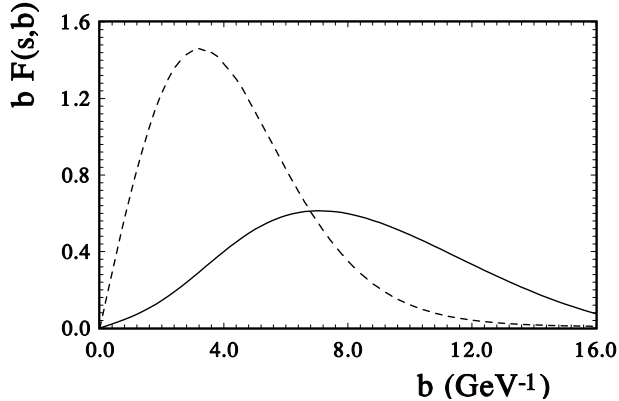} 
\end{center}
\vspace{1.cm}
\caption{ Impact parameter representation of the spin-nonflip and spin-flip amplitude
 at $\sqrt{s}=30$ GeV
($b (1- exp(-\chi_{nf}))$ - dashed line, and ($b (\chi_{sf} \ exp(-\chi_{nf}))$ - hard line)   }
\label{Fig_14}

\end{figure}

\section{Conclusions}

     The Generelized parton distributions (GPDs) make it possible to better understand the fine hadron
     structure and  to obtain the hadron structure in the space frame (impact parameter representations).
     The important property of GPDs consists in that
     they are  tightly connected with the hadron 
      form factors.
           The new HEGS
  model  gives a quantitative     description of elastic nucleon scattering at high energy  with
 a small number of fitting parameters. 
 Our model of the GPDs leads to a good description of the proton and neutron  electromagnetic form factors and their elastic scattering simultaneously.
  A successful description  of the existing experimental data by the model shows that
   the elastic scattering  is determined by the generalized structure of the hadron.
   It allows one to find some new features in the differential cross section of $pp$-scattering in the
   unique experimental data  of the TOTEM collaboration at $ \sqrt{s}=13 $ TeV (small oscillations \cite{Osc13-20}
   and anomalous behavior at small momentum transfer \cite{anom13-20} ).
   The inclusion of the spin-flip parts of the scattering amplitude allows one to describe the low energy
   experimental polarization data of the $pp$ elastic scattering.
     It is shown that the non-perturbative spin-effects at high energies may not be small.

     It should be noted that the real part of the scattering amplitude,
      on which the form 
      of the diffraction dip heavily depends, is
     determined in the framework of the HEGS model only by the complex $\bar{s}$, and hence
      it is tightly connected with the imaginary part of the scattering amplitude
      and satisfies the analyticity and the dispersion relations.
       The HEGS model reproduces well the form and the energy dependence of the diffraction dip  of the proton-proton and proton antiproton
    elastic scattering \cite{Dif-min}.

       The research into the  form and energy dependence of the diffraction minimum of the differential cross sections
    of elastic hadron-hadron scattering at different energies
    will give  valuable information on the structure of the hadron scattering amplitude
     and  hence  the hadron structure and the  dynamics of strong interactions.
    The diffraction minimum 
   is created under a strong impact of the unitarization procedure.
    Its dip depends on the contributions of the real part of the spin-non-flip amplitude and the
    whole contribution of the spin-flip  scattering amplitude.
    In the framework of HEGS model, we show a deep connection between elastic and inealastic cross sections,
    which are tightly connected with the hadron structure at small and large distances.

            Quantitatively, for different thin structures of the scattering amplitude,
          wider  analysis is needed.
          This concerns the fixed intercept taken from the deep inelastic processes and the fixed Regge slope $\alpha^{\prime}$,
           as well  as the form of the spin-flip amplitude.
           Such  analysis requires  a wider range of  experimental data, including the polarization data
        of $A_N(s,t)$, $A_{NN}(s,t)$, $A_{LL}(s,t)$, $A_{SL}(s,t)$.
                        The obtained information on the sizes and energy dependence of the
                        spin-flip and double-flip amplitudes will make it possible
    to better understand 
     the results of  famous experiments carried out by A. Krish at the ZGS  to obtain the
          spin-dependent differential cross sections \cite{Krish1a,Krish1b} and
          the spin correlation parameter $A_{NN}$,   
          and   at the AGS \cite{Krish2} to obtain  the spin correlation parameter $A_{N}$
           showing the significant spin  effects at a large momentum transfer.

       The present analysis, which includes the contributions of the spin-flip amplitudes, also shows a large contradiction between the extracted value of  $\rho(s,t)$ and the predictions from the analysis based on the dispersion relations.
        However, our opinion is that  additional analysis is needed, which will include
     additional corrections connected with the possible oscillation in the scattering amplitude and
    with the $t$-dependence of the spin-flip scattering amplitude.
     We hope that
     future experiments at NICA can give valuable information for the improvement of our theoretical understanding of  strong hadron interaction.

%

\vspace{2.cm}

\end{document}